\documentclass[a4paper,11pt]{article}
\pdfoutput=1 
\usepackage{jheppub} 
\usepackage[utf8]{inputenc}
\usepackage{graphicx}
\usepackage[dvipsnames]{xcolor}
\usepackage{epsfig}
\usepackage{rotating}
\usepackage{amssymb}
\usepackage{subfigure}
\usepackage{dsfont}
\usepackage{psfrag}
\usepackage{amsmath,euscript,array,mathrsfs}
\usepackage{bbold}
\usepackage{epsf}
\usepackage{slashed}

\usepackage[normalem]{ulem}
\usepackage{soul}
\usepackage{cancel}

\makeatletter
\def\@fpheader{%
	\sffamily\small
	Prepared for submission to JHEP\hfill\texttt{IPPP/25/55}%
}
\makeatother

\newcommand{\ignore}[1]{}
\newcommand\SU{\mathrm{SU}}

\newcommand{\beq}{\begin{equation}}
\newcommand{\eeq}{\end{equation}}
\newcommand{\beqs}{\begin{eqnarray}}
\newcommand{\eeqs}{\end{eqnarray}}

\newcommand{\Tr}{{\rm Tr}}

\def\hbar{\hspace{0pt}\raisebox{1pt}{$-$} \hspace{-7pt} h}

\def\ket#1{| #1\rangle}

\newcommand\Emax{E_{\rm max}}
\newcommand\Heff{H_{\rm eff}}

\newcommand\Enfree{\overline{E}_n}

\newcommand{\be}{\begin{equation}}
\newcommand{\ee}{\end{equation}}

\newcommand{\bea}{\begin{eqnarray}}
\newcommand{\eea}{\end{eqnarray}}

\def\lbldef#1#2{\expandafter\gdef\csname #1\endcsname {#2}}

\newcommand{\ber}{\begin{eqnarray}}
\newcommand{\eer}{\end{eqnarray}}

\newcommand{\beqar}{\begin{eqnarray}}

\newcommand{\cE}{{\cal E}}
\newcommand{\cC}{{\cal C}}

\newcommand{\eeqar}{\end{eqnarray}}

\newcommand{\dsl}
  {\kern.06em\hbox{\raise.15ex\hbox{$/$}\kern-.56em\hbox{$\partial$}}}

\newcommand{\eeqarr}{\end{eqnarray}}
\newcommand{\ZZ}{{\rm \kern 0.275em Z \kern -0.92em Z}\;}

\def\CC{{\mathchoice
{\rm C\mkern-8mu\vrule height1.45ex depth-.05ex
width.05em\mkern9mu\kern-.05em}
{\rm C\mkern-8mu\vrule height1.45ex depth-.05ex
width.05em\mkern9mu\kern-.05em}
{\rm C\mkern-8mu\vrule height1ex depth-.07ex
width.035em\mkern9mu\kern-.035em}
{\rm C\mkern-8mu\vrule height.65ex depth-.1ex
width.025em\mkern8mu\kern-.025em}}}

\def\RR{{\rm I\kern-1.6pt {\rm R}}}

\def\ZZ{{\rm Z}\kern-3.8pt {\rm Z} \kern2pt}
\def\IB{\relax{\rm I\kern-.18em B}}
\def\ID{\relax{\rm I\kern-.18em D}}
\def\II{\relax{\rm I\kern-.18em I}}
\def\IP{\relax{\rm I\kern-.18em P}}

\newcommand{\bear}{\begin{eqnarray}}
\newcommand{\eear}{\end{eqnarray}}

\def\to{\rightarrow}

\def\cl{{closed}}

\def\to{\rightarrow}

\def\tt{\tilde{\theta}}

\def\6{\partial}

\def\bea{\begin{eqnarray}}
\def\eea{\end{eqnarray}}

\def\beqx{\begin{displaymath}}
\def\eeqx{\end{displaymath}}

\newcommand{\bmat}{\left(\begin{array}}
\newcommand{\emat}{\end{array}\right)}

\def\ce{{\cal E}}

\def\cl{{\cal L}}

\def\bo{{\raise-.3ex\hbox{\large$\Box$}}}

\def\face{{\raise.2ex\hbox{$\displaystyle \bigodot$}\mskip-2.2mu \llap {$\ddot
        \smile$}}}

\def\>{\rangle}
\def\<{\langle}

\def\leftrightarrowfill{$\mathsurround=0pt \mathord\leftarrow \mkern-6mu
        \cleaders\hbox{$\mkern-2mu \mathord- \mkern-2mu$}\hfill
        \mkern-6mu \mathord\rightarrow$}
\def\dvec#1{\vbox{\ialign{##\crcr
        \leftrightarrowfill\crcr\noalign{\kern-1pt\nointerlineskip}
        $\hfil\displaystyle{#1}\hfil$\crcr}}}

\def\Tr{{\rm Tr \,}}

\def\-{\hphantom{-}}

\title{Hamiltonian Truncation Framework for Gauge Theories on the Interval}

\author[a]{Rachel Houtz}

\affiliation[a]{Department of Physics, University of Florida, Gainesville, FL 32611, USA}

\author[b]{James Ingoldby}

\affiliation[b]{Institute for Particle Physics Phenomenology, Durham University, Durham DH1 3LE, UK}

\date{\today}

\abstract{In this work, we investigate gauge theories in two dimensions nonperturbatively using the Hamiltonian truncation approach. Working on a spatial interval and adopting the axial gauge, we remove all gauge field degrees of freedom and express the interacting Hamiltonian in the eigenbasis of the free Dirac theory, truncated at a finite energy. As a benchmark we analyse the Schwinger model, where our numerical spectra agree closely with the exact results from bosonization across a wide range of couplings, validating the construction of the Hamiltonian. We then generalize the formulation to nonabelian gauge groups and apply it to $\SU(3)$ gauge theory with a single massless Dirac fermion. These results demonstrate that gauge theories can be explored nonperturbatively using a truncated Hamiltonian that generates evolutions in ordinary time, offering a complementary alternative to lattice field theory. 
}

\begin{document}
\maketitle
\flushbottom

\section{Introduction}

Strongly coupled gauge theories remain challenging to study, as quantitative predictions lie beyond the reach of perturbation theory. Analytic methods exploiting the large-$N$ limit or supersymmetry can yield exact solutions or controlled approximations, but cannot be straightforwardly generalized to a wider class of gauge theories. 
Nonperturbative numerical approaches instead offer direct access to more general gauge theories, with lattice gauge theory providing the most established and widely used framework. Here we explore Hamiltonian truncation as a complementary numerical approach for investigating strongly coupled gauge theories.

Hamiltonian truncation is a nonperturbative method that approximates a strongly-coupled quantum theory using a finite-dimensional Hamiltonian matrix~\cite{Brooks:1983sb, Yurov:1989yu, Yurov:1991my}. The infinite basis of states in the Hilbert space is truncated using an energy cutoff, and the resulting effective Hamiltonian can be studied numerically without resorting to a perturbative expansion, with the spectrum obtained by straightforward diagonalization. 
Recent works~\cite{Hogervorst:2014rta,Rychkov:2014eea,Katz:2014uoa} brought renewed attention to the approach and demonstrated its practical value in strongly coupled field theories, and accurate results have been obtained in low dimensions~\cite{Rychkov:2015vap, James:2017cpc, Bajnok:2015bgw, Anand:2017yij, EliasMiro:2021aof, Schmoll:2023eez, Chen:2023glf, Anand:2020qnp, Henning:2022xlj, Fitzpatrick:2022dwq}. This method promises further advances with the development of improvement schemes~\cite{Elias-Miro:2015bqk, Elias-Miro:2017tup,Elias-Miro:2020qwz, Cohen:2021erm, EliasMiro:2022pua,Delouche:2023wsl,Delouche:2024yuo,Demiray:2025zqh} and adaptations to quantum computing~\cite{Liu:2020eoa,Ingoldby:2024fcy, Ingoldby:2025bdb}.

Here we focus on the application of Hamiltonian truncation to strongly coupled gauge theories, where the lattice Monte--Carlo method in Euclidean space has long been the leading numerical tool. Because Hamiltonian truncation starts from the Hamiltonian rather than the Euclidean path-integral formulation, real-time dynamics are directly accessible and the sign problem~\cite{PhysRevLett.46.77,VONDERLINDEN199253,PhysRevE.49.3855} is absent. Moreover, the method can be implemented without introducing a spatial lattice. This is advantageous because lattice discretization necessarily breaks continuous spacetime symmetries such as translations and rotations, and for fermions it triggers the Nielsen–Ninomiya theorem, which obstructs the simultaneous preservation of chiral symmetry and the elimination of fermion doubling~\cite{Wilson:1974sk,Nielsen:1980rz}. By avoiding lattice discretization, Hamiltonian truncation provides a promising approach for investigating fermionic theories while preserving chiral symmetries.

We develop and test a Hamiltonian truncation framework for 1+1D gauge theories, representing an exploratory application of this method to gauge theories in equal-time quantization. The analysis is performed on a finite interval in axial gauge. The downside of this choice is the loss of translational invariance, momentum conservation, and that boundary conditions for fermion fields at the interval endpoints explicitly break chiral symmetry. However, this choice also removes all gauge-field degrees of freedom and allows the theory to be written entirely in terms of matter fields, simplifying the Hamiltonian that forms the start point for truncation. Although our long-term goal is to extend these techniques to richer examples - including higher-dimensional theories of phenomenological interest, the reduced number of degrees of freedom makes these models ideal testbeds. Such simplified settings have already proved valuable in related contexts, including tensor-network simulations~\cite{Banuls:2019bmf,Banuls:2019rao,Montangero:2021puw,Abel:2025pxa,Dempsey:2025wia} and recent proposals for quantum-computing applications to quantum field theory~\cite{Bauer:2022hpo,DiMeglio:2023nsa,Halimeh:2023lid,Halimeh:2025vvp}.

Our analysis begins with 1+1D QED, the Schwinger model~\cite{Schwinger:1962tp}. This simple model has several attractive features. Gauge interactions are strongly relevant in 1+1D, so truncation using an energy cutoff is expected to converge quickly for this simple theory. Secondly, the model exhibits confinement and a mass gap that a successful truncation should reproduce. Finally, the Schwinger model is exactly solvable via bosonization, which provides nonperturbative results for direct comparison~\cite{Coleman:1975pw}. We also extend the same construction to 1+1D nonabelian $\SU(N)$ gauge theory and examine the resulting spectra, finding the appearance of a light, bound and color-singlet meson state.

This paper is organized as follows. In Section~\ref{sec:ht}, we review the basics of Hamiltonian truncation. In Section~\ref{sec:sch}, we introduce the Schwinger model, gauge fix, quantize, and derive its interacting Hamiltonian. We extend this analysis to the $\SU(N)$ gauge theory in Section~\ref{sec:nonabelian}. Our numerical results are presented in Section~\ref{sec:results}, and we end with a discussion of our findings in Section~\ref{sec:disc}.

\section{Hamiltonian Truncation}
\label{sec:ht}

We begin by reviewing the Hamiltonian truncation method.\footnote{For a further review, see~\cite{Fitzpatrick:2022dwq}. } One begins by splitting the Hamiltonian into two parts: 
	\begin{align}
	H	&= H_0 + V \,.
	\end{align}	
The first term $H_0$ is the Hamiltonian of some solved theory with a known energy eigenbasis. A convenient choice for $H_0$ is a free theory, as we employ below. The second term $V$ encodes additional interactions that \it deform \rm the theory away from $H_0$, and can in general include strong interactions. The full $H$ is then written in the known eigenbasis of $H_0$. In order to obtain a finite-sized $H$, we first must discretize the spectrum, for example by working in finite volume. Next, one selects a finite subset of basis states to form the finite Hilbert space using a truncation scheme. Here we truncate using an energy cutoff $E_{\rm max}$. The states in the Hilbert space can be labelled by their known $H_0$ energy eigenvalues:
	\begin{align}
	H_0 \ket{ \Enfree }
		&= \Enfree \ket {\Enfree} \,.
	\end{align}
We retain a subspace of states that satisfy $\Enfree \leq E_{\rm max}$. We use $\Enfree$ to denote energy eigenvalues of $H_0$, and reserve $E_n$ for energy eigenvalues of $H$. The full Hamiltonian can then be approximated by a finite-dimensional $H_{\rm eff}$ that acts on this subspace.

Despite discarding an infinite set of basis states, truncation can perform well when $V$ is governed by relevant operators, meaning operators that flow from weak coupling in the UV to strong coupling in the IR. Intuitively, one expects that the influence of high energy states above $\Emax$ on IR observables is suppressed because relevant operators have decreasing influence at higher energies. Typically, this results in a power law rate of convergence of the spectrum. The power controlling the rate of convergence can be increased  by adding improvement terms~\cite{Elias-Miro:2015bqk, Elias-Miro:2017tup,Elias-Miro:2020qwz, Cohen:2021erm, EliasMiro:2022pua,Delouche:2023wsl,Delouche:2024yuo,Demiray:2025zqh}. In particular, EFT methodology has been used to systematically organize the error corrections in an order-by-order fashion as introduced in~\cite{Cohen:2021erm}. This systematic improvement has been demonstrated to next-to-leading order, at which level nonlocal effects nontrivially conspire to organize themselves into error reduction that improves the rate of convergence from $1/E_{\rm max}^2$ to $1/E_{\rm max}^4$ in 2D $\phi^4$~\cite{Demiray:2025zqh}. The methodology has also been applied to minimal model conformal field theories deformed with relevant operators, which have UV divergences requiring renormalization~\cite{EliasMiro:2022pua,Delouche:2023wsl,Delouche:2024yuo}.

Most truncation studies of gauge theories to date have employed light-cone quantization, yielding major advances in the analysis of 2D gauge dynamics~\cite{Bergknoff:1976xr,Anand:2020gnn, Bhanot:1993xp, Demeterfi:1993rs, Gopakumar:2012gd, Katz:2013qua,Katz:2014uoa,Dempsey:2021xpf, Lunin:1999ib, Fitzpatrick:2019cif,Anand:2021qnd}, and the framework has recently been extended to 3D QED~\cite{Fitzpatrick:2025hqk}. While light-cone quantization is particularly powerful for studying spectra, it offers limited access to vacuum expectation values and spontaneous symmetry breaking~\cite{Fitzpatrick:2023aqm}. Equal-time Hamiltonian truncation\footnote{Here, equal-time means that the field theory is quantized by imposing commutation or anticommutation relations between fields with the same time coordinate $t$.} provides a natural complement: it enables the direct study of vacuum structure, real-time dynamics, and facilitates comparison with lattice Hamiltonian results. Related progress has also been made applying variational Hamiltonian methods to lattice gauge theories, that rely on local truncations of the gauge degrees of freedom~\cite{Fontana:2024rux}, offering an alternative route to exploring gauge theories nonperturbatively.

\section{The Schwinger Model}
\label{sec:sch}

QED in 1+1 dimensions, known as the Schwinger model, has the following Lagrangian
\begin{align}\label{eq:L}
	\cl = -\frac{1}{4}F_{\mu\nu}F^{\mu\nu}+\frac{g\theta}{4\pi}\epsilon_{\mu\nu}F^{\mu\nu}+\bar{\psi}\left(i\slashed{\partial}-g\slashed{A}-m\right)\psi\,,
\end{align}
where we label the coordinates $x^0=t$ and $x^1=x$. We use the metric convention is $\eta^{\mu\nu}=\text{diag}(1,-1)$, and take $\epsilon_{01}=-\epsilon^{01}=1$. We also use the following conventions for the Dirac matrices, $\gamma^0=\sigma_3$, $\gamma^1=i\sigma_2$ and $\gamma^5=\gamma^0\gamma^1$. In 1+1 dimensions, the only independent, nonvanishing component of the electromagnetic field tensor is $F_{01}\equiv\partial_0 A_1 - \partial_1 A_0$, where $A_{\mu}$ is the gauge potential. $F_{01}$ corresponds to the electric field, while there is no magnetic field in 1+1 dimensions.

The gauge coupling, $g$, has dimensions of mass. As a result, processes with characteristic energies much larger than $g$ can be treated perturbatively, while the dynamics becomes strongly coupled and nonperturbative in the infrared. Remarkably, for massless fermions the Schwinger model is exactly solvable: it describes a theory of noninteracting pseudoscalar mesons. The theta term above can be interpreted as a constant background electric field of strength $\cE_B=g\theta/2\pi$~\cite{coleman1976more}.

\subsection{Classical Hamiltonian}

We now turn to the task of constructing a Hamiltonian for the interval Schwinger model which is suitable for nonperturbative analysis using Hamiltonian truncation. In $d=1+1$ there are no transverse directions, so the electromagnetic field carries no propagating degrees of freedom. On a finite interval, it is a consistent simplification to work in axial gauge
\begin{align}\label{eq:axial}
	A_1 = 0\,,
\end{align}
which eliminates the spatial component of the gauge field, while leaving only $A_0$ as a nondynamical variable without a kinetic term. This choice is free of pathologies~\cite{PhysRev.127.1821,Weinberg:1995mt}, and is also used to simplify lattice Hamiltonian formulations of gauge theories on the interval~\cite{Farrell:2022wyt,Ciavarella:2023mfc}. 

The axial gauge Hamiltonian on the interval of length $L$ can then be found from Eq.~(\ref{eq:L}) using
\begin{align}\label{eq:hschem}
	H = \int_0^L dx \; \Pi_\psi \dot{\psi} - \cl\,,
\end{align}
where $\Pi_\psi\equiv\frac{\partial\cl}{\partial\dot{\psi}}=i\bar{\psi}\gamma^0$ is the momentum conjugate of the fermion field. The momentum conjugate for the $A_0$ field vanishes, since it is nondynamical.

We can obtain a simpler expression for the interval Hamiltonian, in which all the gauge fields are fully eliminated, by using the constraint equation for $A_0$ that comes from extremizing the Schwinger model action. In general, this equation will include both bulk and boundary terms. In order for the variational principle to be well defined, these boundary terms must vanish, which requires appropriate boundary conditions for the $A_0$ field.

A consistent choice is to set the electric field, $\ce\equiv F_{01}$, to be equal to its background value indicated by $\theta$ at the left boundary of the interval, while setting the $A_0$ field to a constant at the right boundary
\begin{align}\label{eq:bcgauge}
	\left.\partial_1 A_0\right|_{x=0} = - \ce(x=0) = - \frac{g\theta}{2\pi}\,, && \left.A_0\right|_{x=L} = 0\,,
\end{align}
which leaves the electric field at the right boundary undetermined. Physically, fixing $\ce(0)=g\theta/2\pi$ can be viewed as introducing a background charge localized at the left boundary that sources the electric field, while the condition at $x=L$ corresponds to grounding the right boundary. With this choice, the $A_0$ field satisfies
\begin{align}\label{eq:gauss}
	\partial_1^2 A_0 + g j^0 = 0\,,
\end{align}
where $j^\mu\equiv\bar{\psi}\gamma^\mu\psi$ is the electric current density. This constraint equation is none other than Gauss' law in axial gauge.

The electric field is then fully determined in terms of the fermion fields. Integrating Eq.~(\ref{eq:gauss}) and applying the boundary conditions yields
\begin{align}\label{eq:efield}
	\ce(x,t)  =  g\int_0^xdy\,j^0(y,t) + \frac{g\theta}{2\pi}\,.
\end{align}
The electric field at the right boundary is then determined by the field at the left boundary, and the total electric charge within the interval.  In this picture the right boundary can absorb any additional charge carried in the bulk, so that Gauss' law relates the bulk charge to the difference of electric fields at the two ends.

In addition to the gauge field boundary conditions discussed above, we must also specify boundary conditions for the fermion fields. Writing the two-component Dirac spinor as
\begin{align}
	\psi = \begin{pmatrix}
		\psi_u \\ \psi_d
	\end{pmatrix},
\end{align}
we note that in the basis of Dirac matrices we are using the components $\psi_{u,d}$ are convenient for formulating boundary conditions, although they are linear combinations of the left and right-moving chiral modes.

On the interval, the fermion boundary conditions must remove the boundary contributions from the action, ensuring a consistent variational problem, while preserving the gauged $U(1)$ vector symmetry $\psi \to e^{i\alpha}\psi$. By contrast, they explicitly break the global axial symmetry $\psi \to e^{i\alpha\gamma^5}\psi$. The admissible conditions fall into two classes, directly analogous to the Ramond (R) and Neveu--Schwarz (NS) cases on the circle, corresponding respectively to periodic and anti--periodic boundary conditions for the fermion field.

For the \emph{Ramond} class, the full set of inequivalent possibilities (excluding field redefinitions of the type $\psi\rightarrow\gamma^5\psi$) is \cite{Okuda:2022hsq}
\begin{align}
	\psi_d(x=0)=\psi_d(x=L)=0 \,, \qquad
	\text{or equivalently}\qquad
	\psi_u(x=0)=\psi_u(x=L)=0 \,,
\end{align}
whereas for the \emph{Neveu--Schwarz} class, the full set of inequivalent possibilities is
\begin{align}
	\psi_d(x=0)=0 \,, \qquad \psi_u(x=L)=0 \,,
	\label{eq:ns}
\end{align}
together with the variant obtained by exchanging $u \leftrightarrow d$.

It is worth noting that these interval boundary conditions have a direct analogue in the Kogut-Susskind staggered lattice formulation: Ramond-type boundary conditions correspond to taking an odd number of staggered lattice sites, while Neveu--Schwarz-type conditions correspond to an even number of staggered sites~\cite{Okuda:2022hsq}. More generally, the two classes lead to different low-energy fermionic spectra once the theory is quantized, so it is essential to fix the boundary conditions at the outset before proceeding with Hamiltonian truncation.

Starting from Eq.~(\ref{eq:hschem}), and simplifying it using Gauss' law and the boundary condition in Eq.~(\ref{eq:bcgauge}) allows us to derive a simple expression for the Hamiltonian that forms our start point for quantization
\begin{align}\label{eq:ham}
	H = \int_0^L dx \,\bar{\psi}(-i\gamma^1\partial_1+m)\psi+\frac{\ce^2}{2}\,,
\end{align}
where the electric field $\ce$ is expressed entirely in terms of fermionic fields using Eq.~(\ref{eq:efield}). We see that the gauge potentials $A_\mu$ can be completely eliminated from the Hamiltonian on the interval.

\subsection{Quantization}

We quantize the Schwinger model canonically, by promoting the fermion fields to operators, and imposing canonical anticommutation relations
\begin{gather}
	\{\psi_\alpha(x,t),\,\psi^\dagger_\beta(y,t)\}_+ = \delta_{\alpha\beta}\delta(x-y)\,, \nonumber \\	\{\psi_\alpha(x,t),\,\psi_\beta(y,t)\}_+ = 	\{\psi^\dagger_\alpha(x,t),\,\psi^\dagger_\beta(y,t)\}_+ =0\,,
	\label{eq:can-comms}
\end{gather}
between fields at equal times. Here, $\alpha$ and $\beta$ index over the two components $u$ or $d$ of the fermion fields. Setting the time coordinate set to zero for simplicity, the fermion field operator has the following mode expansion
\begin{align}
	\psi(x) = \sum_{n=0}^{\infty}a_nu_n(x)+b^\dagger_nv_n(x)\,,
	\label{eq:psi}
\end{align} 
where the $u_n$ and $v_n$ are the complete set of particle and antiparticle solutions to the Dirac equation for a free massive fermion (without the gauge interaction) on the interval. Here $a_n$ denotes an annihilation operator for fermions and $b_n^\dagger$ a creation operator for antifermions. They are given the standard anticommutation relations
\begin{align}\label{eq:comms}
	\{a_n,\,a^\dagger_m\}_+ = \delta_{nm}\,,&&	\{a_n,\,a_m\}_+ = 0\,, &&	\{a_n,\,b_m\}_+ = 0\,, \nonumber \\
	\{b_n,\,b^\dagger_m\}_+ = \delta_{nm}\,,&&	\{b_n,\,b_m\}_+ = 0\,, &&	\{a_n,\,b^\dagger_m\}_+ = 0\,,
\end{align}
which ensure that Eq.~(\ref{eq:can-comms}) is satisfied.

In the discussion that follows, and in all of our numerical results, we adopt the Neveu-Schwarz-type boundary condition indicated in Eq.~(\ref{eq:ns}). The corresponding solutions to the Dirac equation are
\begin{align}
	u_n(x) = \frac{1}{\sqrt{\omega_nL}}\left( \begin{array} {c}
		\sqrt{\omega_n+m}\,\cos(k_nx)	\\ 	i\sqrt{\omega_n-m}\, \sin(k_nx)
	\end{array}\right)\,,
	\label{eq:un}
\end{align}
with $k_n=\pi(n+\frac{1}{2})/L$ for all nonnegative integers $n=0,\,1,\,2,\dots$ The corresponding mode energy is given by $\omega_n=\sqrt{k^2_n+m^2}$. We note that this boundary condition choice removes all zero energy modes, even in the $m=0$ theory, simplifying our analysis. We also have the antiparticle solutions
\begin{align}
	v_n(x) = \frac{1}{\sqrt{\omega_nL}}\left( \begin{array} {c}
		\sqrt{\omega_n-m}\,\cos(k_nx)	\\ -i	\sqrt{\omega_n+m}\, \sin(k_nx)
	\end{array}\right)\,.
	\label{eq:vn}
\end{align}

The classical Hamiltonian in Eq.~(\ref{eq:ham}) can be written as the sum of a solvable Hamiltonian (that of a free Dirac fermion) and an interaction term, that goes as electric field squared. As discussed in Section~\ref{sec:ht}, this decomposition into a solvable piece, $H_0$, and an interaction, $V$, is the starting point for Hamiltonian truncation. After quantization, the free and solvable part of the Hamiltonian takes the form
\begin{align}\label{eq:h0}
	H_0	&= \sum_{n=0}^\infty\omega_n \left( a_n^\dagger a_n + b^\dagger_n b_n\right)\,,
\end{align}
where a constant term, which contributes to the vacuum energy but has no dynamical effect, has been set to zero.

The Hilbert space of the theory is spanned by the eigenstates of $H_0$, which are the Fock states
\begin{align}\label{eq:basisstates}
	\ket{\Enfree} =\prod_{i=0}^{\infty}\left(a^\dagger_i\right)^{r_i}\prod_{j=0}^{\infty}\left(b^\dagger_j\right)^{\bar{r}_j} \ket{0}\,,
\end{align}
where states are labeled with two lists of occupation numbers: $r_j$ gives the occupation state of the $j$th fermion mode (0 or 1), while $\bar{r}_j$ describes the state of the $j$th antifermion mode. Note that the bar in $\ket{\Enfree}$ is not related to antiparticles, but instead is used to differentiate the eigenbasis of $H_0$ from the eigenbasis $\ket{E_n}$ of the interacting Hamiltonian used below. Ordering all antifermion operators to the right of all fermion operators in the definition of the states represents our convention choice.

The total electric charge is a conserved quantity. Its operator in the quantum theory is
\begin{align}
	\mathcal{Q}	&= \sum_{n=0}^\infty \left( a_n^\dagger a^{\phantom{\dagger}}_n - b^\dagger_n b^{\phantom{\dagger}}_n \right)\,,	
\end{align}
so that the basis states from Eq.~(\ref{eq:basisstates}) are also eigenstates of total electric charge
\begin{align}
	\mathcal{Q} 	\ket{\Enfree}	 &= Q 	\ket{\Enfree} \,,
	\qquad Q	= \sum_{n=0}^\infty \left(r_n - \bar r_n\right) \,.
	\label{eq:Qdef}
\end{align}
Charge conservation also ensures that our gauge interaction $V$ does not mix states of different charge $Q$.

We now turn to constructing the interaction $V$ as an operator in the quantum theory. To begin, we require that the electric field must be a good physical operator with finite matrix elements between Fock states of the form in Eq.~(\ref{eq:basisstates}). We construct it by first inputting the field operator from Eq.~(\ref{eq:psi}) into the classical definition from Eq.~(\ref{eq:efield}). After setting the $\theta$ parameter to zero for simplicity, we find
\begin{align}\label{eq:efieldquantum}
		:\ce(x): = g\sum_{n,m=0}^\infty\left(e^{ff}_{nm}(x)a^\dagger_na_m+e^{f\bar{f}}_{nm}(x)a^\dagger_nb^\dagger_m+e^{\bar{f}f}_{nm}(x)b_na_m-e^{\bar{f}\bar{f}}_{nm}(x)b^\dagger_mb_n\right)\,,
\end{align}
where $f$ and $\bar{f}$ refer to the fermion and antifermion terms from the sum in Eq.~(\ref{eq:psi}). If we introduce an index $\alpha \in \{ f, \bar f \}$, then all the tensors $e^{\alpha_1,\alpha_2}_{nm}(x)$ may be compactly defined as integrals over the mode functions from Eqs.~(\ref{eq:un}) and (\ref{eq:vn}) through
\begin{align}
	e^{\alpha_1\alpha_2}_{nm}(x) \equiv \int_0^x dy\, w_{\alpha_1\,n}^\dagger(y)\,w_{\alpha_2\,m}^{\phantom{\dagger}}(y)\,,
\end{align}
where $w_{\alpha\,n}(x)$ represents the solution $u_n(x)$ if $\alpha=f$, or the solution $v_n(x)$ if $\alpha=\bar{f}$. In Eq.~(\ref{eq:efieldquantum}), the dots in $:\ce(x):$ have been included to emphasize that in defining the operator in the quantum theory, we have chosen to \emph{normal order} by moving all creation operators to the left of all annihilation operators, and setting any additive constants that arise from using the anticommutation relations to zero. Normal ordering this way ensures the required finiteness of the matrix elements. It is also responsible for the sign appearing in the last term of Eq.~(\ref{eq:efieldquantum}). 

Promoting Gauss' Law in Eq.~(\ref{eq:gauss}) to an operator equation in the quantum theory, as we have done in Eq.~(\ref{eq:efieldquantum}), also implicitly fixes a renormalization condition: using a non-gauge-invariant regulator could in principle introduce an extra term $m_\gamma^2 A_0$ in the quantum version of Eq.~(\ref{eq:gauss}), but our scheme corresponds to choosing $m_\gamma=0$. This point is significant because the one-loop photon self-energy is superficially UV divergent in $1+1$ dimensions, and its cancellation is not manifest when using a non-gauge-invariant regulator such as the $E_{\rm max}$ truncation.

With the electric field well defined, we are finally able to construct $V$ as an operator using
\begin{align}\label{eq:vquantum}
	V = \frac{1}{2}\int_0^Ldx\,\left(:\ce(x):\right)^2\,.
\end{align}
This form for $V$ is a natural choice, given its classical predecessor Eq.~(\ref{eq:ham}). However, it means that $V$ itself is not normal ordered. Nevertheless, in Section~\ref{sec:results} we provide strong numerical evidence that it is the \emph{correct} choice, by computing the spectrum nonperturbatively and showing that it agrees with exact analytic results derived from bosonization. 

By combing Eqs.~(\ref{eq:efieldquantum}) and (\ref{eq:vquantum}), we can compactly express $V$ as a sum of four-body terms with calculable coefficients
\begin{align}\label{eq:intschem}
	V = \frac{g^2L}{2}\sum_{\alpha_i \in \{f,\bar f\}}
	\;\sum_{n,m,k,l=0}^{\infty}\, 
	V^{\alpha_1,\alpha_2,\alpha_3,\alpha_4}_{nmkl}\, 
	:c^\dagger_{\alpha_1,n}c^{\phantom{\dagger}}_{\alpha_2,m}:\,
	:c^\dagger_{\alpha_3,k}c^{\phantom{\dagger}}_{\alpha_4,l}:\,,
\end{align}
where $c_{\alpha\,n}$ represents the operator $a_n$ if $\alpha=f$, or the operator $b^\dagger_n$ if $\alpha=\bar{f}$. The first two operators in each term should be normal ordered amongst themselves, and so should the last two. However, each term is not required to be normal ordered overall. The coefficients are defined by
\begin{align}
	 V^{\alpha_1,\alpha_2,\alpha_3,\alpha_4}_{nmkl} \equiv \frac{1}{L}\int_0^L dx \left[\int_0^xdy\,w^\dagger_{1,n}(y)\,w^{\phantom{\dagger}}_{2,m}(y)\right]\left[\int_0^xdy\,w^\dagger_{3,k}(y)\,w^{\phantom{\dagger}}_{4,l}(y)\right]\,.
	 \label{eq:vcoeffs1}
\end{align}
The prefactor of $1/L$ ensures that these coefficients are dimensionless.

Finally, to simplify the numerical evaluation of matrix elements, we recast $V$ as a sum of fully normal ordered terms. By applying the commutation relations from Eq.~(\ref{eq:comms}), the interaction can be written as
\begin{align}\label{eq:VSchwinger}
	V 	&=\frac{g^2L}2 \sum_{n,m,k, l=0}^\infty
	\left[
	\kappa_{nmkl}^{(1)} a^\dagger_n a^\dagger_m b^\dagger_k b^\dagger_l
	+\kappa_{nmkl}^{(2)} \left(a^\dagger_n a^\dagger_m b^\dagger_k a_l 
	+ b^\dagger_n b^\dagger_m a^\dagger_k b_l\right)
	\right. \nonumber\\
	&\qquad \qquad \qquad \quad  \left.
	+ \kappa_{nmkl}^{(3)} \left(a^\dagger_n a^\dagger_m a_k a_l + b^\dagger_n b^\dagger_m b_k b_l\right)
	+ \kappa_{nmkl}^{(4)} a^\dagger_n b^\dagger_m b_k a_l
	\right]
	\nonumber\\[8pt]
	&\qquad
	+\frac{g^2L}2 \sum_{n,l=0}^\infty \left[
	\kappa_{nl}^{(5)} a^\dagger_n b^\dagger_l
	+ \kappa_{nl}^{(6)}\left( a^\dagger_n a_l
	+ b^\dagger_n b_l\right)
	\right] + \text{h.c.,}
\end{align}
where explicit formulae for the dimensionless tensors $\kappa^{(i)}$ are provided in appendix~\ref{app:tensors}. We note that we have left out a term proportional to the identity operator, which has an infinite prefactor. This corresponds to a renormalization of the vacuum energy, and has no effect on energy differences, or on dynamics. All the terms shown above have finite prefactors. We note also that $V$ is invariant under charge conjugation symmetry $\cC a_n \cC^{-1} = b_n,\;\; \cC b_n \cC^{-1} = a_n$.

The equation (\ref{eq:VSchwinger}) for $V$ is used to derive all of our numerical results for the Schwinger model in Section~\ref{sec:results}. To numerically build the explicit truncated Hamiltonian for this analysis, we compute matrix elements of $V$ between the Fock states shown in Eq.~(\ref{eq:basisstates}). To facilitate this, we make use of the following Jordan-Wigner transformation
\begin{alignat}{3}
	a^\dagger_i &= \prod_{k=0}^{i-1}\!\left(-\sigma_k^z\right)\sigma_i^+,&\quad
	a_i &= \prod_{k=0}^{i-1}\!\left(-\sigma_k^z\right)\sigma_i^-,&\quad
	a^\dagger_i a_i &= \tfrac{1}{2}\!\left(\mathbf{1} + \sigma_i^z\right), \nonumber\\
	b^\dagger_j &= \prod_{k=0}^{j+N_m-1}\!\left(-\sigma_k^z\right)\sigma_{j+N_m}^+,&\quad
	b_j &= \prod_{k=0}^{j+N_m-1}\!\left(-\sigma_k^z\right)\sigma_{j+N_m}^-,&\quad
	b^\dagger_j b_j &= \tfrac{1}{2}\!\left(\mathbf{1} + \sigma_{j+N_m}^z\right),
	\label{eq:jw}
\end{alignat}
where $N_m$ denotes the highest occupied fermion mode included in the truncated Fock basis, with antifermion operators labeled relative to this maximum. The Jordan-Wigner transformation maps fermionic operators, which anticommute across modes, to spin operators that commute when acting on different modes. This serves as a bookkeeping device for tracking the minus signs that appear when bringing operators into the conventional ordering used to define the states in Eq.~(\ref{eq:basisstates}).

\section{Nonabelian Gauge Theory}
\label{sec:nonabelian}

In this section, we apply Hamiltonian truncation to the $\SU(N)$ gauge theory on the interval, with one Dirac fermion in the fundamental representation of the gauge group. Fortunately, many of the same logical steps that were used to formulate the Schwinger model as a Hamiltonian truncation problem can be generalized to the $\SU(N)$ case. We start from the Lagrangian density
\begin{align}
	\cl \;=\; -\frac{1}{2}\,\Tr\!\left[F_{\mu\nu}F^{\mu\nu}\right]
	+ \bar{\psi}\,\left(i\slashed{D}-m\right)\psi \,,
\end{align}
where $F_{\mu\nu} = \partial_\mu A_\nu - \partial_\nu A_\mu - ig\,[A_\mu,A_\nu]$ is the nonabelian field strength, $A_\mu = A_\mu^a T^a$ with generators $T^a$ in the fundamental of $\SU(N)$ normalized as $\Tr(T^a T^b) = \tfrac{1}{2}\delta^{ab}$, and $D_\mu = \partial_\mu - ig A_\mu$ is the covariant derivative. 

In contrast to the Schwinger model, there is no independent $\theta$ parameter in 1+1 dimensions for $\SU(N)$, since the gauge invariant generalization of the corresponding Lagrangian term $\epsilon_{\mu\nu}\Tr{\left[ F^{\mu\nu}\right]}$ vanishes identically. The dynamics are therefore fully specified by the gauge coupling $g$, the fermion mass $m$, and the interval length $L$.

Again, we derive the Hamiltonian using axial gauge $A_1=0$, which is also a consistent choice in the nonabelian case~\cite{PhysRev.127.1821,Weinberg:1995mt}. It also has the virtue of removing the gluon self interaction term in $F_{\mu\nu}$. The boundary conditions we choose for the gauge fields are straightforward generalizations of those we adopted for the Schwinger case: $\partial_1 A_0|_{x=0}=0$ and $A_0|_{x=L}=0$. With this choice, the chromoelectric fields are set to zero at the left boundary but can consistently be nonzero at the right. We will make use of this feature in our numerical analysis.

The classical Hamiltonian for the $\SU(N)$ theory then takes the form
\begin{align}
	H = \int_0^L dx \,\sum_{j=1}^N\bar{\psi}_j(-i\gamma^1\partial_1+m)\psi_j+\sum_{a=1}^{N^2-1}\frac{\left(\ce^{a}\right)^2}{2}\,,
\end{align}
where for concreteness we have now included and explicitly summed over the fundamental gauge index $j$ carried by the fermion field, and the adjoint index $a$ carried by the chromoelectric field, $\ce^a$. This quantity is expressed in terms of the fermion fields through an analogous equation to Eq.~(\ref{eq:efield}), except with the current replaced by its nonabelian analog $j^{a\,0}=\bar{\psi}T^a\gamma^0\psi$, and with the $\theta$ term removed.

After quantization, the solvable part of the Hamiltonian takes the form
\begin{align}
	H_0 = \sum_{j=1}^{N}\sum_{n=0}^\infty \omega_n \left( a_{n,j}^\dagger a_{n,j} + b_{n,j}^\dagger b_{n,j} \right)\,,
	\label{eq:h0naren}
\end{align}
with independent fermion and antifermion creation and annihilation operators $a_{n,j}^\dagger, a_{n,j}$ and $b_{n,j}^\dagger, b_{n,j}$ assigned to each color component $j$ of the fundamental representation. By convention, we represent the eigenstates of Eq.~(\ref{eq:h0naren}) as
\begin{align}
	\ket{\Enfree} = \prod_{j=1}^N\left[\prod_{k=0}^{\infty}\left(a^\dagger_{k,j}\right)^{r_{k,j}}\prod_{l=0}^{\infty}\left(b^\dagger_{l,j}\right)^{\bar{r}_{l,j}}\right] \ket{0}\,,
	\label{eq:basisna}
\end{align}
so that all the operators that create particles of fundamental color index 1 come to the left of all the color 2 operators, which are to the left of the color 3 operators, e.t.c.

The interaction term in the Hamiltonian generalizes from Eq.~(\ref{eq:intschem}) as follows
\begin{align}
	V = \frac{g^2L}{2}\sum_{\alpha_i \in \{f,\bar f\}}\, \sum_{n,m,p,q=0}^{\infty}\,V^{\alpha_1,\alpha_2,\alpha_3,\alpha_4}_{nmpq} \,\sum_{a=1}^{N^2-1}\sum_{i,j,k,l=1}^{N} :c^\dagger_{\alpha_1,n,i}T^a_{ij}c^{\phantom{\dagger}}_{\alpha_2,m,j}:\,:c^\dagger_{\alpha_3,p,k}T^a_{kl}c^{\phantom{\dagger}}_{\alpha_4,q,l}:\,,
	\label{eq:intnaschem}
\end{align}
where helpfully the same tensors $V^{\alpha_1,\alpha_2,\alpha_3,\alpha_4}_{nmkl}$ that appear in the Schwinger model Hamiltonian reappear here. They are defined in Eq.~(\ref{eq:vcoeffs1}). The sum over adjoint gauge group indices can be done by hand using
\begin{align}
	\tau_{ijkl} \equiv \sum_{a=1}^{N^2-1}T^a_{ij}T^a_{kl} = \frac{1}{2}\left(\delta_{il}\delta_{jk}-\frac{1}{N}\delta_{ij}\delta_{kl}\right)\,.
	\label{eq:taus}
\end{align}
Moreover, the tensor $\tau_{ijkl}$ defined above enables us to write the interaction more compactly.

Just as in the Schwinger case, the interaction can be recast into a fully normal ordered form. The nonabelian generalization of Eq.~(\ref{eq:VSchwinger}) is
\begin{align}
	V 	&=\frac{g^2L}2 \sum_{n,m,p, q} \sum_{ijkl}
	\left[
	\kappa_{nmpq}^{(1)} \tau_{ikjl} 
	a^\dagger_{n, i} a^\dagger_{m, j} b^\dagger_{p, k} b^\dagger_{q, l}
	+\kappa_{nmpq}^{(2)} \tau_{iljk}\left(
	a^\dagger_{n, i} a^\dagger_{m, j} b^\dagger_{p, k} a_{q, l} + \text{ch.c}
	\right)
	\right. \nonumber\\
	&\quad  \left.
	+ \kappa_{nmpq}^{(3)} \tau_{ikjl} \left(
	a^\dagger_{n, i} a^\dagger_{m, j} a_{k, p} a_{l, q} + \text{ch.c} \right)
	+ \left( \kappa_{nmpq}^{(4a)} \tau_{ijkl} + \kappa_{nmpq}^{(4b)} \tau_{ilkj} \right)
	a^\dagger_{n,i} b^\dagger_{m, j} b_{p, k} a_{q, l}
	\right]
	\nonumber\\[8pt]
	&\ \ \ 
	+\frac{g^2L}2 \ \sum_{nq} \sum_{il} \left[
	\kappa_{nq}^{(5)} C_2(\Box) \delta_{il}
	a^\dagger_{n, i} b^\dagger_{q, l}
	+ \kappa_{nq}^{(6)}  C_2(\Box) \delta_{il}
	\left(a^\dagger_{n, i} a_{q, l}+\text{ch.c}\right)
	\right]
	+ \text{h.c.,}
	\label{eq:VsuN}
\end{align}
where the $\kappa^{(i)}$ tensors are given in appendix~\ref{app:tensors}, and the color tensors $\tau$ are defined in Eq.~(\ref{eq:taus}). All terms in the last line are proportional to the quadratic casimir of the fundamental representation  $C_2(\Box)=\left(N^2-1\right)/(2N)$. Just as in the Schwinger case, we drop a divergent contribution to $V$ proportional to the identity operator, which has no effect on energy differences or dynamics. Finally, we use the notation ch.c to represent the charge conjugate. This is the term you obtain by performing the swaps $a_{n,i}\leftrightarrow b_{n,i}$ and $a^\dagger_{n,i}\leftrightarrow b^\dagger_{n,i}$. This is the form of the interaction for $\SU(N)$ gauge theory that we will use for Hamiltonian truncation in Section~\ref{sec:results}.

To properly keep track of signs when acting with creation and annihilation operators on the basis states of the form shown in Eq.~(\ref{eq:basisna}), we also make use of a Jordan-Wigner transform to spin operators, analogous to Eq.~(\ref{eq:jw}). The crucial difference in the $\SU(N)$ case though is that we have $N$ types of creation and $N$ types of annihilation operators per mode. If the maximum number of any occupied mode in the truncated basis is $N_m$, we need $2NN_m$ spin operators, rather than just $2N_m$ in the Schwinger case. The spins should be grouped as indicated in Eq.~(\ref{eq:basisna}): the first \(N_m\) spins encode the fermion modes of color~1, the next \(N_m\) encode the corresponding antifermion modes, followed by the fermion and antifermion blocks for color~2, and so on through all \(N\) colors.

We can reduce the size of our truncated basis by imposing global symmetry constraints from the outset. In the nonabelian case, the constraints we use are conservation of baryon number and conservation of a subset of the nonabelian charges. The baryon number operator is
\begin{align}\label{eq:baryon}
	\mathcal{B} = \frac{1}{N}\sum_{i=1}^{N}\sum_{n=0}^\infty\left( a_{n,i}^\dagger a_{n,i} - b^\dagger_{n,i} b_{n,i} \right)\,,
\end{align}
while the nonabelian charge operators are
\begin{align}\label{eq:nonabQ}
	\mathcal{Q}^a = \sum_{i,j=1}^{N}\sum_{n=0}^\infty\left( a_{n,i}^\dagger T^a_{ij}a_{n,j} - b^\dagger_{n,i}T^a_{ij} b_{n,j} \right)\,.
\end{align}

In the present work we restrict attention to the sector of the Hilbert space with vanishing baryon number, $\mathcal{B}\ket{\psi}=0$. In the large-volume limit only states satisfying $\mathcal{Q}^a\ket{\psi}=0$ for all generators $T^a$ survive in the spectrum, while configurations carrying nonzero nonabelian charges acquire energies of order $g$ and correspond to finite-volume excitations with charges pinned to the right boundary~\cite{Farrell:2022wyt}.
Ideally one would therefore impose $\mathcal{Q}^a\ket{\psi}=0$ for all $a$ when constructing the truncated Hilbert space, but in the occupation-number representation used here, the exact singlet states are complicated linear combinations of basis vectors. 

To balance Hilbert-space reduction with ease of matrix-element computation, we retain only the charge constraints generated by the diagonal (Cartan) elements of $\SU(N)$. These abelian charges are straightforward to enforce directly on occupation numbers, significantly reducing the Hilbert-space dimension while retaining a structured basis. The price to be paid is that the truncated spectrum will contain spurious states with nonzero off-diagonal charges and energies $\mathcal{O}(g)$, which can be identified and discarded at the end. Our choice of boundary conditions for the gauge field, which did not require the chromoelectric fields at the right boundary to vanish, allow us to include these extra $\SU(N)$ nonsinglet states consistently.

\section{Results}
\label{sec:results}

Here we present the numerical results from the truncated Hamiltonian formulation of both the abelian Schwinger model and nonabelian $\SU(3)$ gauge theory in 1+1D on the interval. We start with the Schwinger model, which has the advantage of having a bosonized version to check our results against. 

The Hamiltonian we study here can be written as a sum of the free theory and interaction term, $H=H_0 + V$, where $H_0$ is given in Eq.~(\ref{eq:h0}) and $V$ in Eq.~(\ref{eq:VSchwinger}). For simplicity, we will set both the fermion mass and the background electric field to zero (i.e. $m=\theta=0$). Each term in $V$ can be written as a matrix in the truncated basis of eigenstates of $H_0$, which take the form shown in Eq.~(\ref{eq:basisstates}). We truncate by some $\Emax$ such that we only keep states with $\Enfree \leq \Emax$. After truncation, our finite-dimensional, effective Hamiltonian acting on the truncated basis is denoted by $\Heff$. The code to construct and diagonalize $\Heff$ was written in \texttt{Python}, and results obtained on a laptop computer.\footnote{The software that we used for both the Schwinger model and the nonabelian gauge theory is available at \href{https://github.com/rahoutz/HT_gauge_interval}{\texttt{github.com/rahoutz/HT\_gauge\_interval}} and \href{https://github.com/james-ingoldby/HT-2d-GaugeTheory}{\texttt{github.com/james-ingoldby/HT-2d-GaugeTheory}}.}

Note that with the fermion mass set to zero, $1/L$ is an overall prefactor in $\Enfree$ and appears only in the combination $g^2L$ in $V$, and so one is free to rescale the effective Hamiltonian by changing both $L$ and $g$, while keeping their product, $gL$, fixed. To facilitate comparison and highlight the underlying behavior independent of rescaling, all the plots below use dimensionless quantities multiplied by the appropriate factors of $L$, such as $\Emax L$, $gL$, and so on. For efficiency, in results presented here we have further restricted the basis to a subset of states with $Q=0$, see Eq.~(\ref{eq:Qdef}), as charge conservation ensures the gauge interaction in $V$ does not mix states of different charge $Q$.

The massless Schwinger model can be equivalently expressed as the theory of one free scalar field of mass $M^2_S\equiv g^2/\pi$, as a result of bosonization~\cite{Coleman:1975pw}. The scalar field must satisfy boundary conditions on the interval. If we demand that the scalar field vanishes at the interval endpoints, the energy levels of the theory are given by
	\begin{align}
		E = \sum_{j=1}^\infty n_j \Omega_j\,, \qquad \Omega_j = \sqrt{\left(\frac{j\pi}{L}\right)^2 + M^2_S}\,,
		\label{eq:int-boson}
	\end{align}
where $\Omega_j$ gives the energy of each scalar field mode $j$ which satisfies the boundary conditions and the $ n_j \in \mathbb{Z}_{\ge 0} $ are the corresponding occupation numbers.

As discussed in Sec.~\ref{sec:ht}, convergence is predicted only for relevant operators, and in the Schwinger model the gauge interaction in $V$ is indeed relevant. This implies that its effect weakens at high energies and becomes perturbative for sufficiently large $\Emax$, which underlies the EFT-inspired improvement program of~\cite{Cohen:2021erm}. We can estimate from a simple power-counting argument that the leading truncation error in the spectrum of $H$ should scale as
	\begin{align}
	\delta E	&\sim \frac{g^4 L }{E_{\rm max}^{2}} \,,
	\label{eq:nda}
	\end{align}
where the factor of $L$ comes from the spatial integral, $g^4$ from the next order in the coupling expansion, and simple dimensional analysis sets $1/\Emax^2$. For this reason, many of our convergence plots are presented as a function of $1/\Emax^2$ such that the predicted scaling appears as a straight line.

\begin{figure}[h!]
\centering
	\begin{minipage}{.9\textwidth}
	\centering
		\includegraphics[height=0.4\columnwidth, trim = .2cm .2cm 0cm .2cm, clip=true]{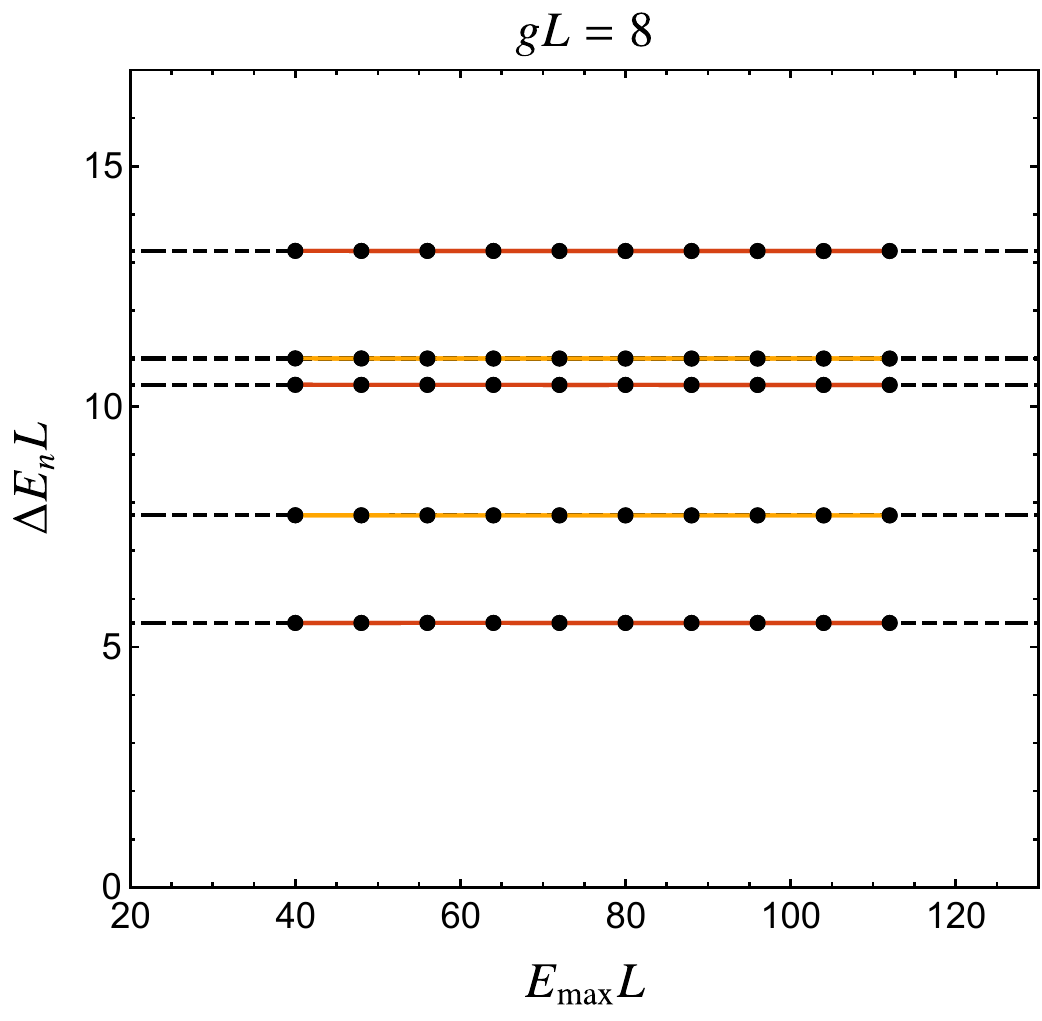} \qquad 
		\includegraphics[height=0.4\columnwidth, trim = .2cm .2cm 0cm .2cm, clip = true]{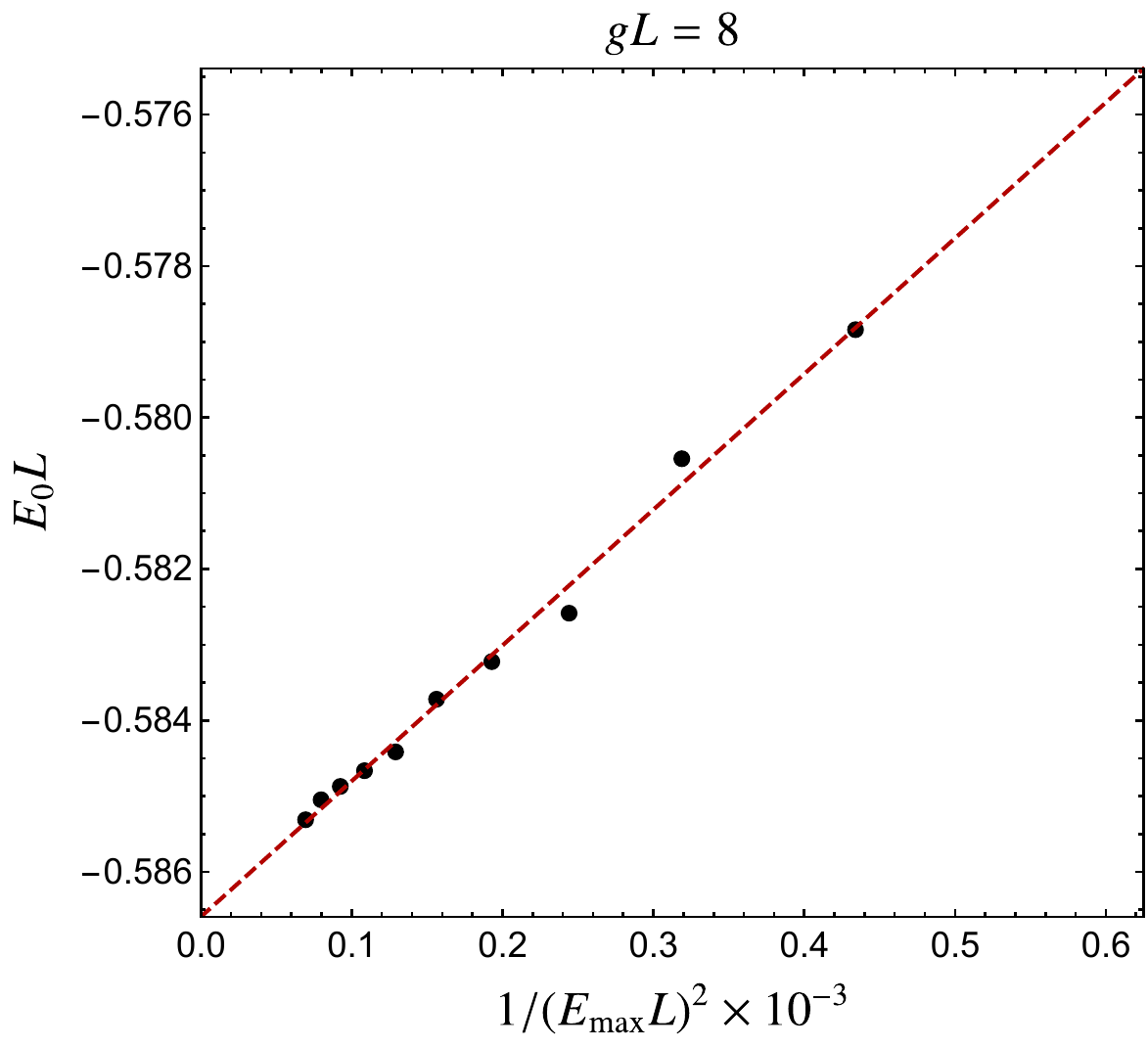}
	\caption{The left plot shows the excited energy gaps $\Delta E_n L$ of the Schwinger model  at moderate coupling $gL = 8$. The result from truncation is denoted by dots and connected by red or orange lines. The exact result from bosonization is given in black dashed lines. The right plot shows vacuum energy $E_0 L$ plotted against $1/(\Emax L)^2$ in black dots, along with its linear fit given by a dashed red line. }
	\label{fig:schg1}
	\end{minipage}
\end{figure}
\begin{figure}[h!]
\centering
	\begin{minipage}{.9\textwidth}
	\centering
		\includegraphics[width=.32\textwidth, trim=.2cm 0cm 0cm 0cm, clip=true]{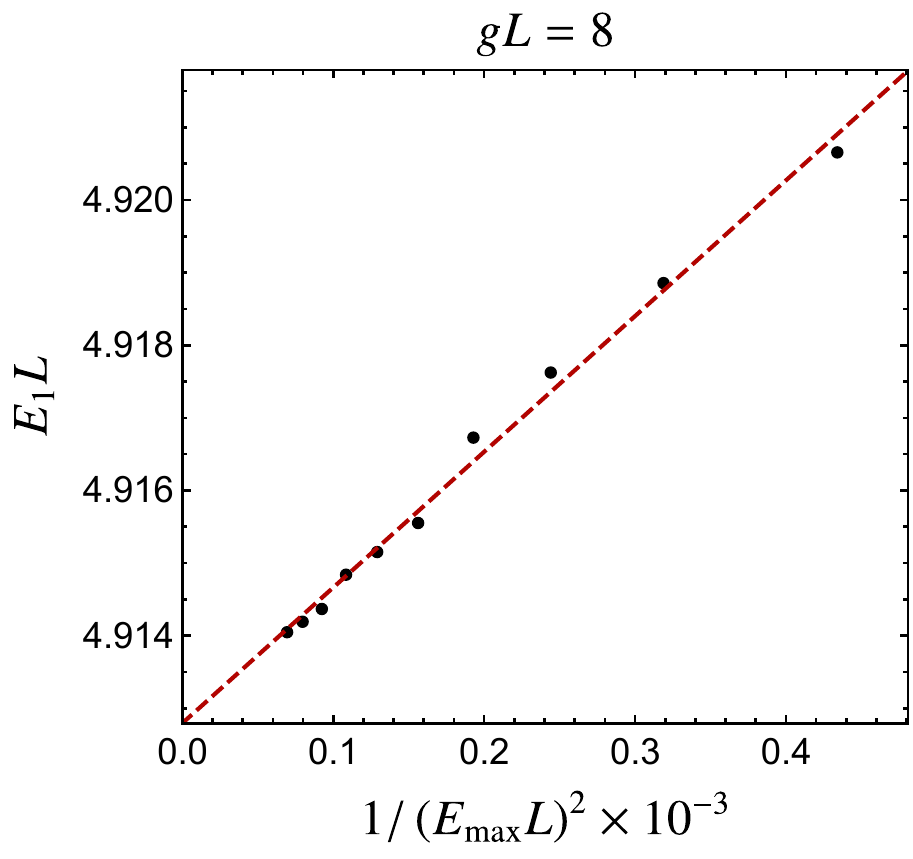}
		\hspace{.001\textwidth}
		\includegraphics[width=.32\textwidth, trim=.2cm 0cm 0cm 0cm, clip=true]{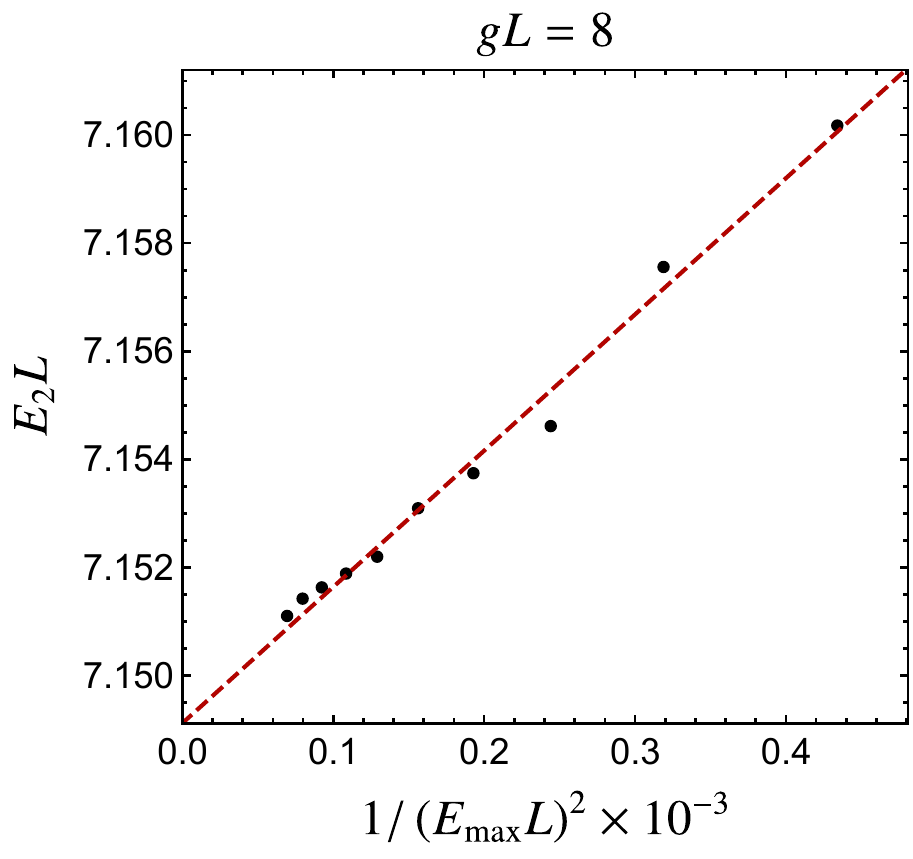}
		\hspace{.001\textwidth}
		\includegraphics[width=.32\textwidth, trim=.2cm 0cm 0cm 0cm, clip=true]{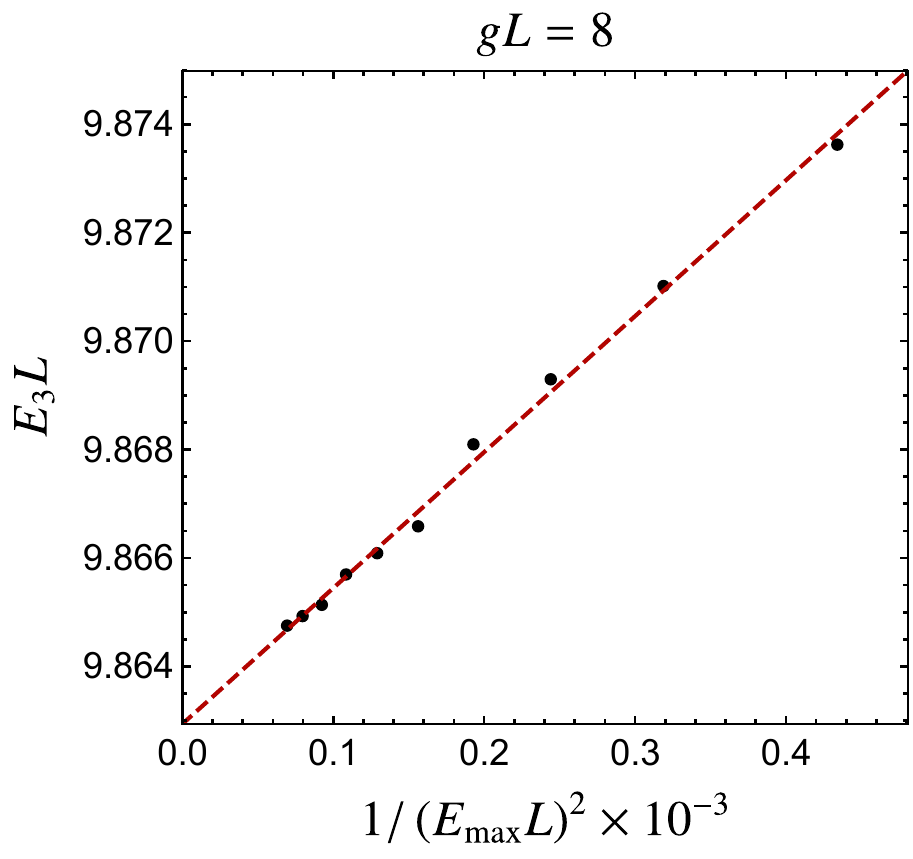}
	\caption{The convergence of the first few excited states in the spectrum for the Schwinger model at moderate coupling $gL = 8$. The energy eigenvalues $E_n L$ are shown versus $1/(\Emax L)^2$. Values obtained from truncation are denoted by black dots, and a linear fits of the energy levels versus $1/(\Emax L)^2$ are shown as red dashed lines. The expected scaling with $\Emax L$ holds for the low-lying eigenstates. }
	\label{fig:schg1_excited}
    	\end{minipage}
\end{figure}

We show the data at moderate coupling, $gL=8$ in Fig.~\ref{fig:schg1}. The size of the basis for each $\Emax L$ is given in Table~\ref{tab:sch-basis}. The plot on the left shows the excitation spectrum above the ground state, 
	\begin{align}
	\Delta E_n = E_n - E_0 \,,
	\label{eq:DE}
	\end{align}
where $E_n$ are the energy eigenvalues of $H_{\rm eff}$. The energy gaps are in close agreement with the results from bosonization, which are indicated by black horizontal lines. The rate of convergence is shown in the right plot, and agree with the expectation of power law convergence $\sim1/\Emax^2$ (for fixed $g, L$). We also show the convergence of the excited energy levels $E_n$ for $n=1,2, 3$, which continues to follow $1/\Emax^2$ scaling, see Fig.~\ref{fig:schg1_excited}. 

\begin{table}[h!]
\centering
	\begin{minipage}{.9\textwidth}
	\centering
	\begin{tabular}{|c|c|}
	\hline
	$\Emax L$	& \# of states in basis	\\
	\hline
	40		&	272	\\
	80		&	9,296	\\
	120		&	146,785	\\
	\hline
	\end{tabular}
	\caption{The number of states in the Hilbert space of the Schwinger model on the interval for a few benchmark values of $\Emax L$.}
	\label{tab:sch-basis}
	\end{minipage}
\end{table}

\begin{figure}[h!]
\centering
	\begin{minipage}{.9\textwidth}
	\centering
		\includegraphics[height=0.4\columnwidth]{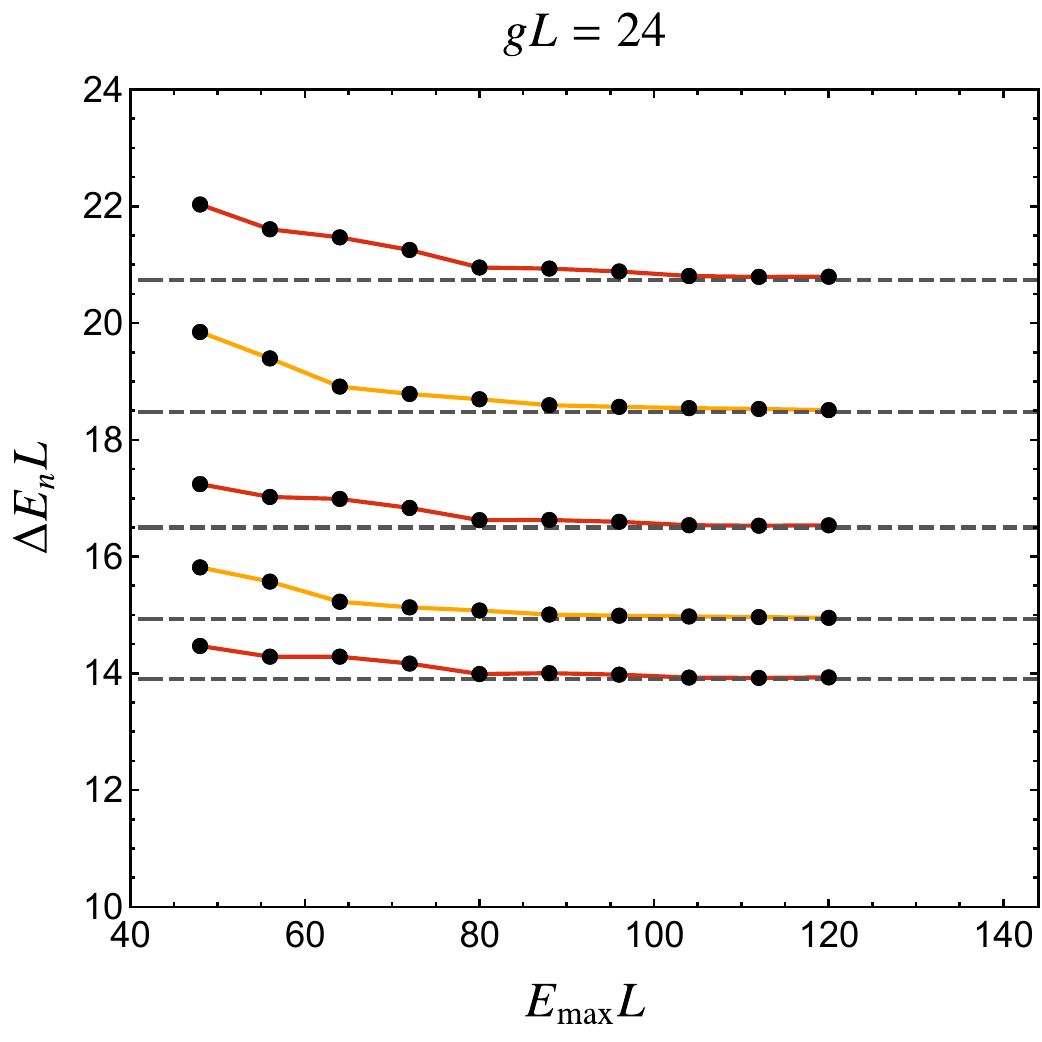} \qquad 
		\includegraphics[height=0.4\columnwidth]{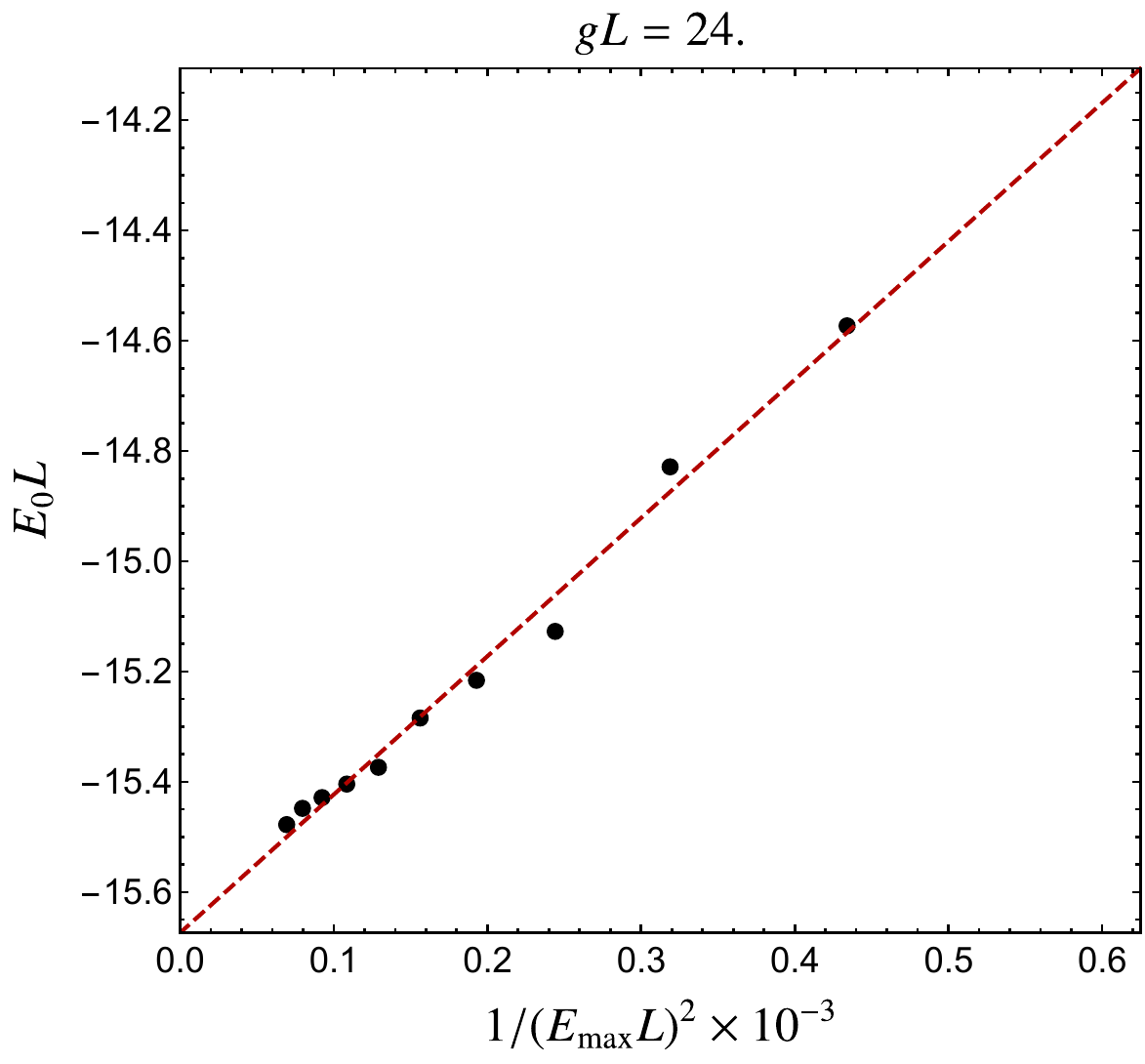}
	\caption{The left plot shows the excited energy gaps $\Delta E_n L$ of the Schwinger model at strong coupling $gL = 24$. The result from truncation is denoted by dots and connected by colored lines. The exact result from bosonization is given in black dashed lines. The numerical data converges towards the exact result. The right plot shows the $1/(\Emax L)^2$ convergence of the vacuum energy $E_0 L$ as $\Emax L$ is increased.}
	\label{fig:schg3}
	\end{minipage}
\end{figure}
\begin{figure}[h!]
	\centering
	\begin{minipage}{.9\textwidth}
	\centering
		\includegraphics[width=.32\textwidth, trim=.2cm .2cm 0cm 0cm, clip=true]{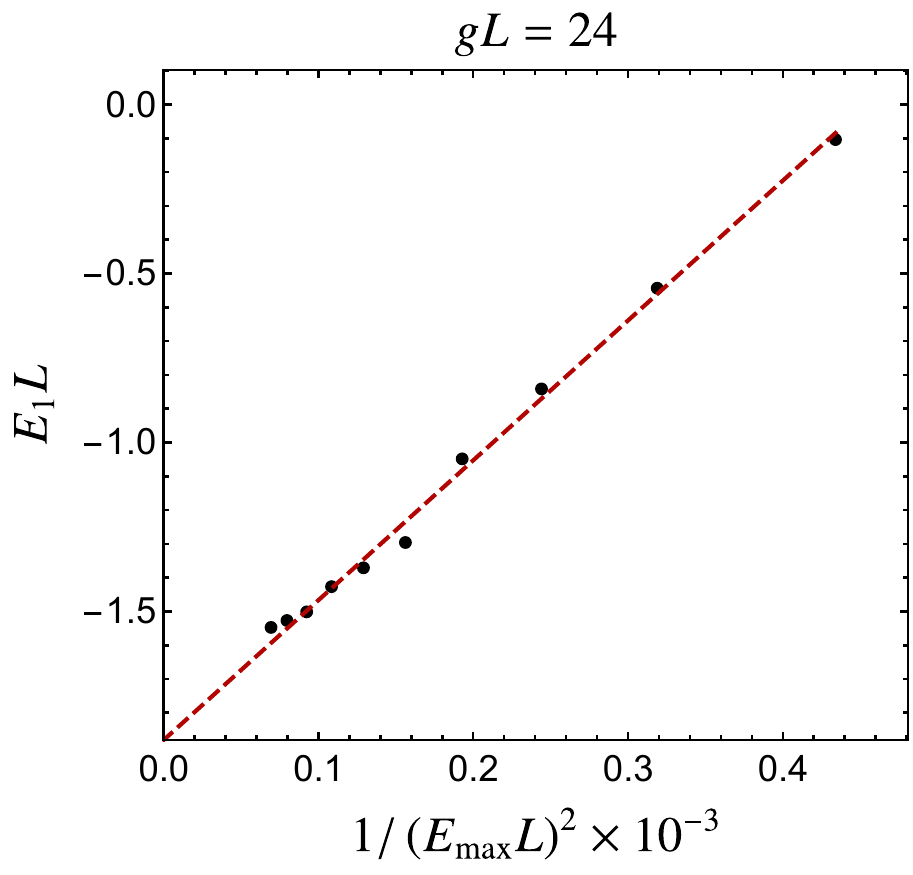}
		\hspace{.001\textwidth}
		\includegraphics[width=.32\textwidth, trim=.2cm .2cm 0cm 0cm, clip=true]{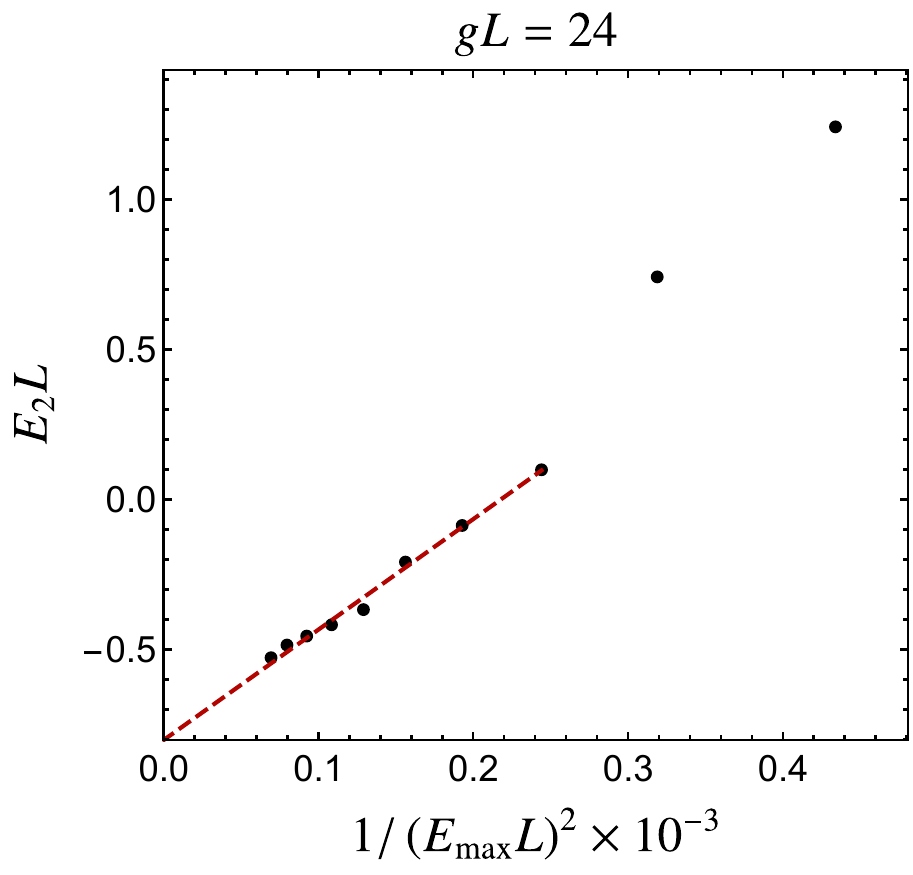}
		\hspace{.001\textwidth}
		\includegraphics[width=.32\textwidth, trim=.2cm .2cm 0cm 0cm, clip=true]{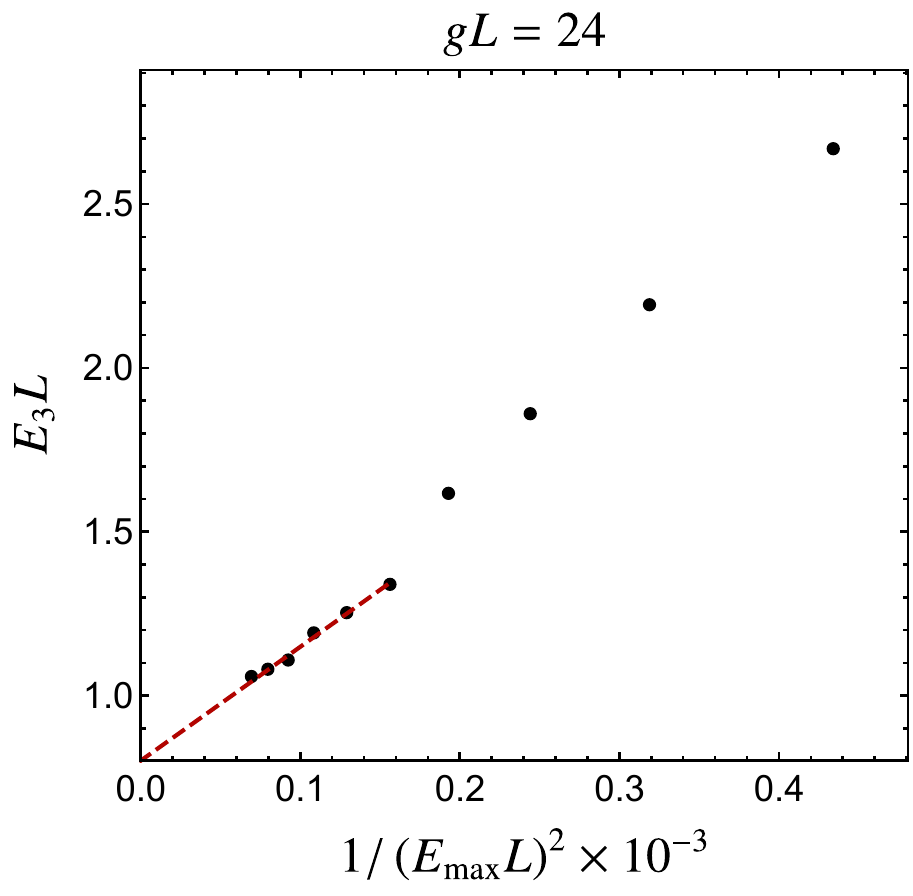}
	\caption{The convergence of the first few excited states in the spectrum for the Schwinger model at moderate coupling $gL = 24$. The energy eigenvalues $E_n L$ are shown versus $1/(\Emax L)^2$. Values obtained from truncation are denoted by black dots, and the $1/(\Emax L)^2$ fits are shown as red dashed lines. The domain of the fit lines indicate the regions over which the fits were performed. Here we see that for higher excited states, the expected scaling emerges only in the high $\Emax L$ tails.}
	\label{fig:schg3_excited}
	\end{minipage}
\end{figure}

Next, we show data for the Schwinger model at strong coupling in Fig.~\ref{fig:schg3}. Even deep into the nonperturbative regime, the truncation performs well. The excited energy gaps approach their exact values from bosonization as $\Emax$ is increased, shown on the left plot in Fig.~\ref{fig:schg3}. The convergence of the vacuum energy $E_0$ also continues to scale with $1/\Emax^2$ as shown on the right. We also show the convergence of the excited energy levels $E_n$ for $n=1,2, 3$. In this case, the power law scaling only emerges in the high $\Emax$ tails, see Fig.~\ref{fig:schg3_excited}.

\begin{figure}[h!]
\centering
	\begin{minipage}{.9\textwidth}
	\centering
		\includegraphics[width=.48\textwidth, trim=.2cm .2cm 0cm 0cm, clip=true]{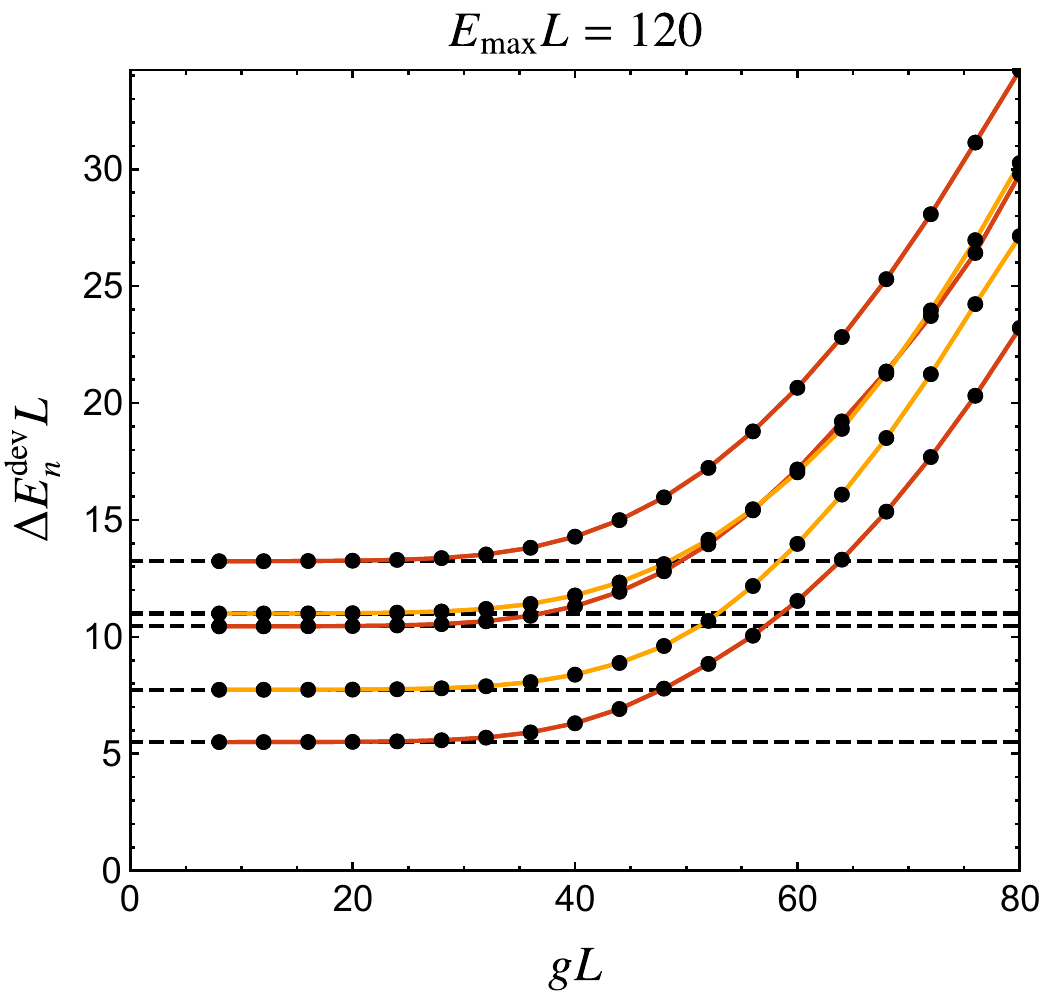}
		\hspace{.01\textwidth}
		\includegraphics[width=.48\textwidth, trim=.2cm .2cm 0cm 0cm, clip=true]{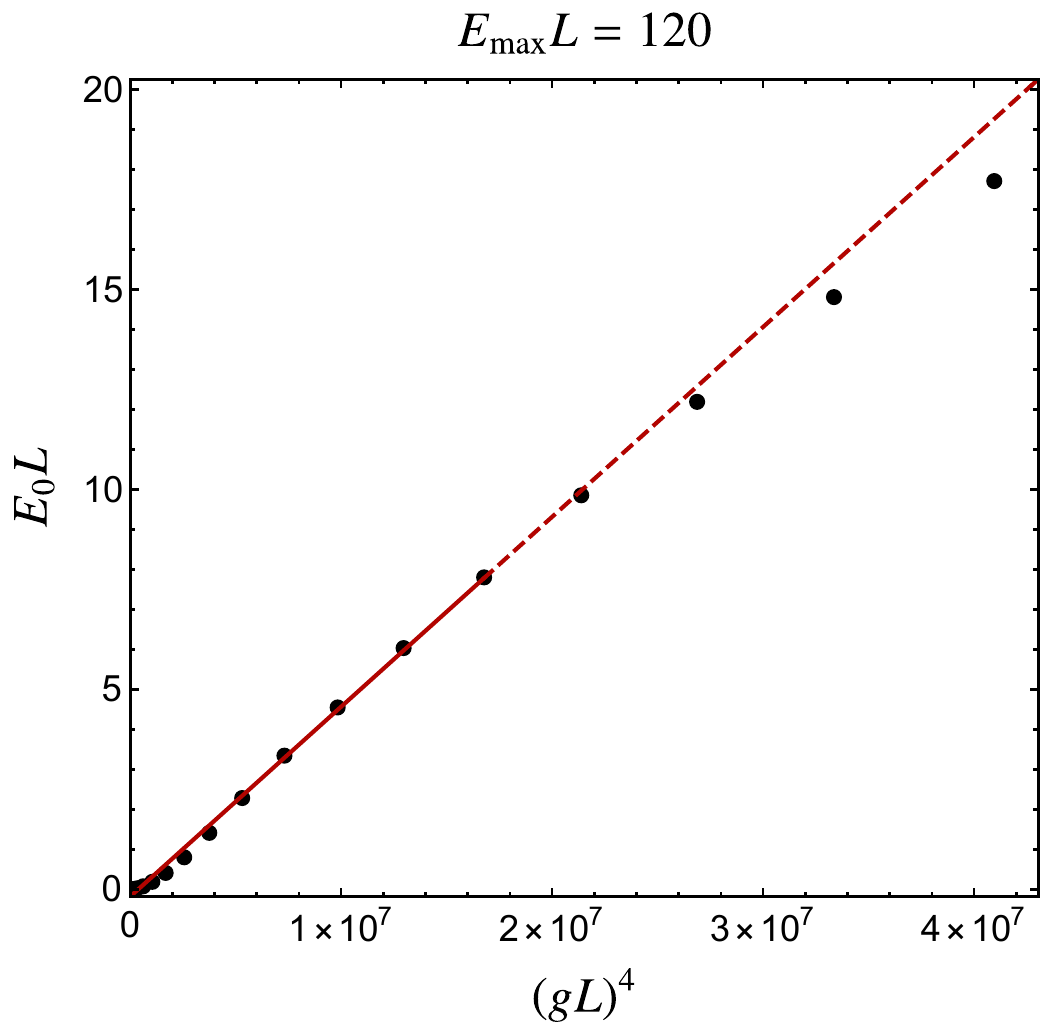}
	\vspace{-5pt}
	\caption{The spectrum as $gL$ is varied. On the left hand side, we plot the deviation in energy gaps of $H_{\rm eff}$ from the predicted bosonized result. To guide the eye, we show the bosonized result from $gL =8$ with black dashed lines. We see that as $gL$ grows, the agreement with the exact bosonized result deteriorates. On the right hand side, we plot the spectrum against $(gL)^4$. We demonstrate its scaling using a linear fit, shown in red. Only points with the smallest $gL$ values are used in the fit, indicated by the range of the solid red line, while the dashed red line shows the fit's extrapolation to larger $gL$ values.}
	\label{fig:g4-scale}
	\end{minipage}
\end{figure}

We also examine the convergence of the spectrum as a function of the gauge coupling $g$ in Fig.~\ref{fig:g4-scale}.  On the left plot, we show how the results from truncation deviate from the bosonization prediction as the coupling $g$ increases. The quantity plotted is:
	\begin{align}
	\Delta E_n^{\rm dev}(g) \equiv \Delta E_n (g) - \Delta  E_n^{\rm bos} (g) +  \Delta E_n^{\rm bos} (gL = 8) 
	\label{eq:devEn}
	\end{align}
where $\Delta E_n$ is the energy gap given in Eq.~(\ref{eq:DE}) and $\Delta E_n^{\rm bos}$ is the analogous quantity extracted from bosonization. The last term is included for visual clarity: it offsets each excited state's energy gap deviation by the $\Delta E_n^{\rm bos}$ result at moderate coupling, $gL = 8$, so that the spectrum is vertically separated. On the right plot we show the convergence of the spectrum with varied coupling. According to the scaling relation estimated in Eq.~(\ref{eq:nda}), the spectrum should converge like $g^4$ holding $L, \Emax$ fixed. We perform a linear fit to the $E_0 L$ vs. $(gL)^4$ data, but omit points with $gL>64$, as at such strong couplings there is noticeable deviation from the expected power-law scaling, likely due to higher order $gL$ effects. For a large range of couplings, the $(gL)^4$ power law scaling closely matches the data from our truncated Hamiltonian. We have therefore independently demonstrated both the $g$ and $\Emax$ power law convergence of the spectrum. This also suggests that there is no need for an improvement term at $\mathcal{O}(g^2)$ that would be analogous to that found in the staggered fermion lattice approach~\cite{Dempsey:2022nys}.

Finally, we turn to the nonabelian $\SU(3)$ gauge theory. In this case, our Hamiltonian is again formed as $H = H_0 + V$, where now $H_0$ is given in Eq.~(\ref{eq:h0naren}) and $V$ is given in Eq.~(\ref{eq:VsuN}). We again choose the basis of $H_{\rm eff}$ using an energy cutoff, and restricting to a subspace where the baryon number as defined in Eq.~(\ref{eq:baryon}) is zero, and where the \emph{diagonal} nonabelian charges are required to vanish. See the discussion at the end of Section~\ref{sec:nonabelian}. The size of the basis for a given $\Emax L$ is given in Table~\ref{tab:su3-basis}. 

\begin{figure}[h!]
\centering
	\begin{minipage}{.9\textwidth}
		\centering
		\includegraphics[height=0.46\textwidth]{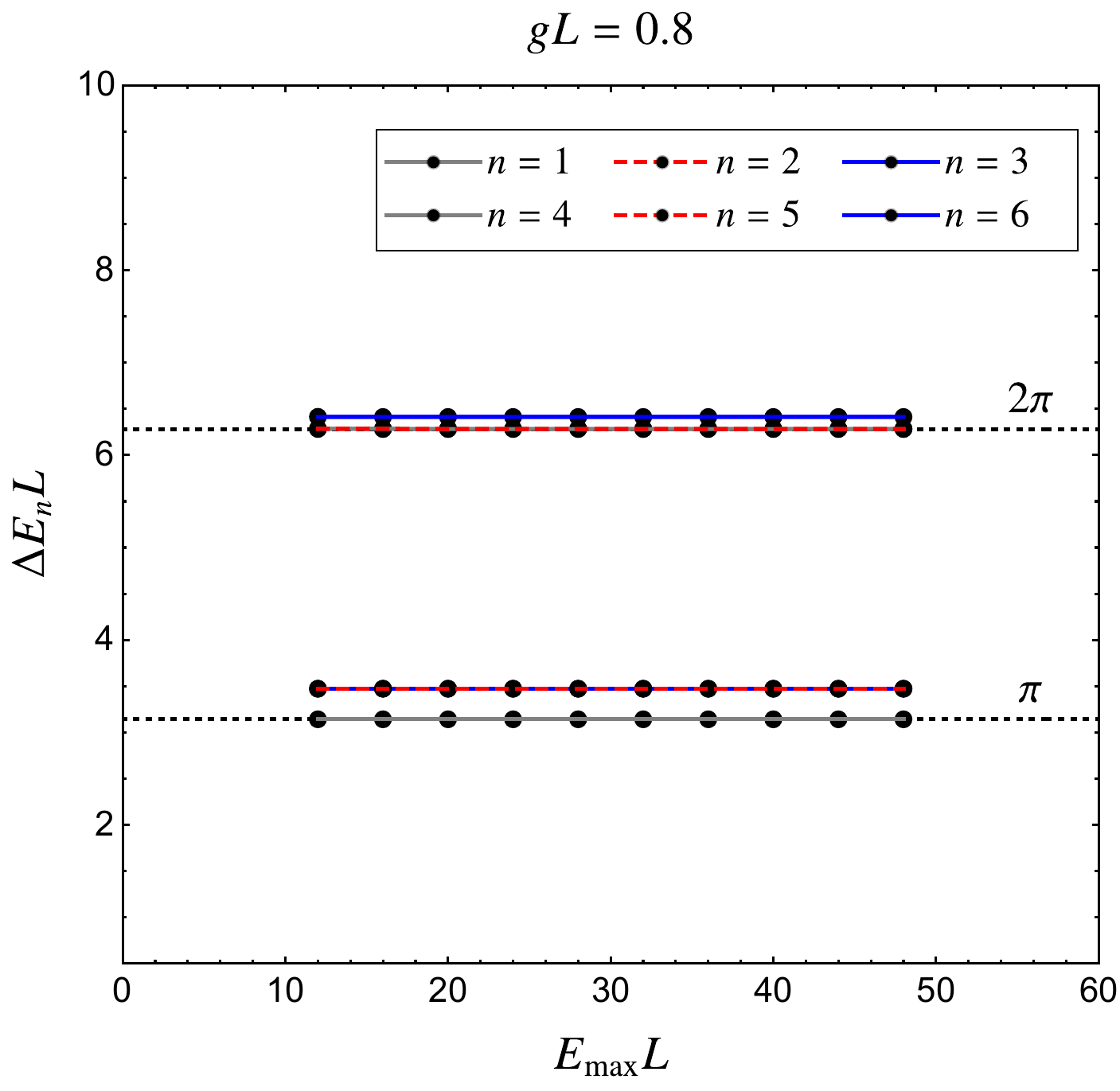} 
		\hspace{.001\textwidth}
		\includegraphics[height=0.46\textwidth]{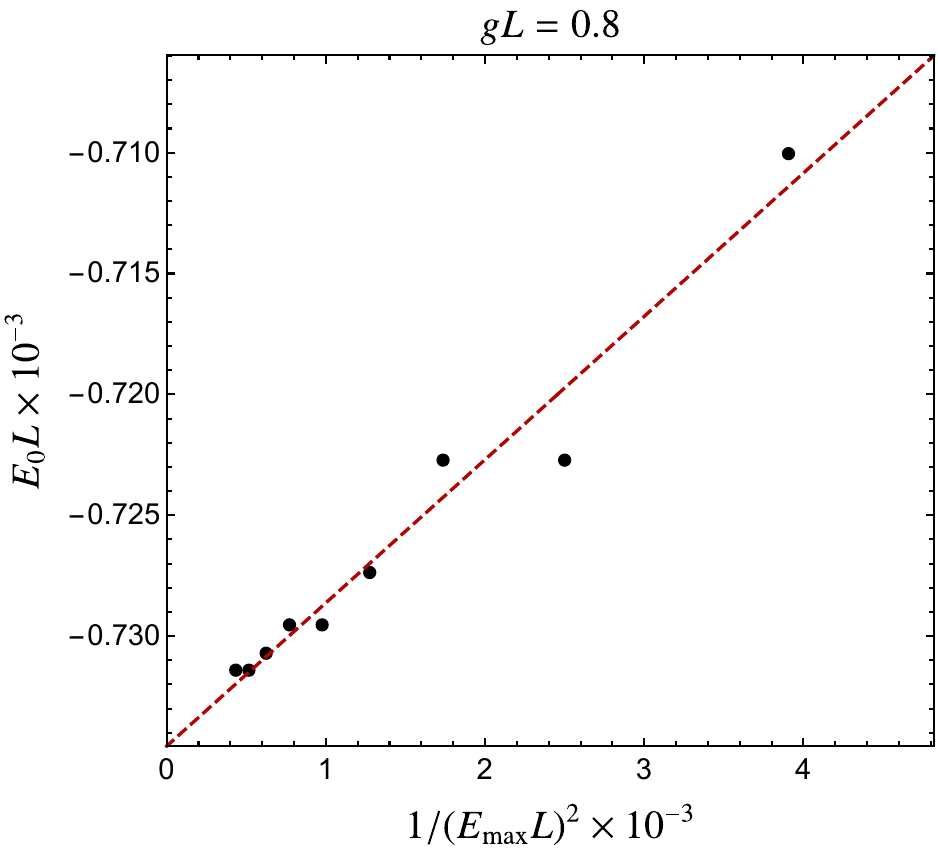}
	\caption{
	The left plot shows the excited energy gaps $\Delta E_n L$ of the $\SU(3)$ model at weak coupling $gL = 0.8$. The result from truncation is denoted by dots and connected by colored lines. The first two energy gaps of the free theory are shown in dotted black  lines. The right plot shows vacuum energy $E_0 L$ plotted against $1/(\Emax L)^2$ in black dots, along with its linear fit given by a dashed red line.
	}
	\label{fig:su3g1}
	\end{minipage}
\end{figure}

\begin{figure}[h!]
	\centering
	\begin{minipage}{.9\textwidth}
	\centering
		\includegraphics[height=0.46\columnwidth]{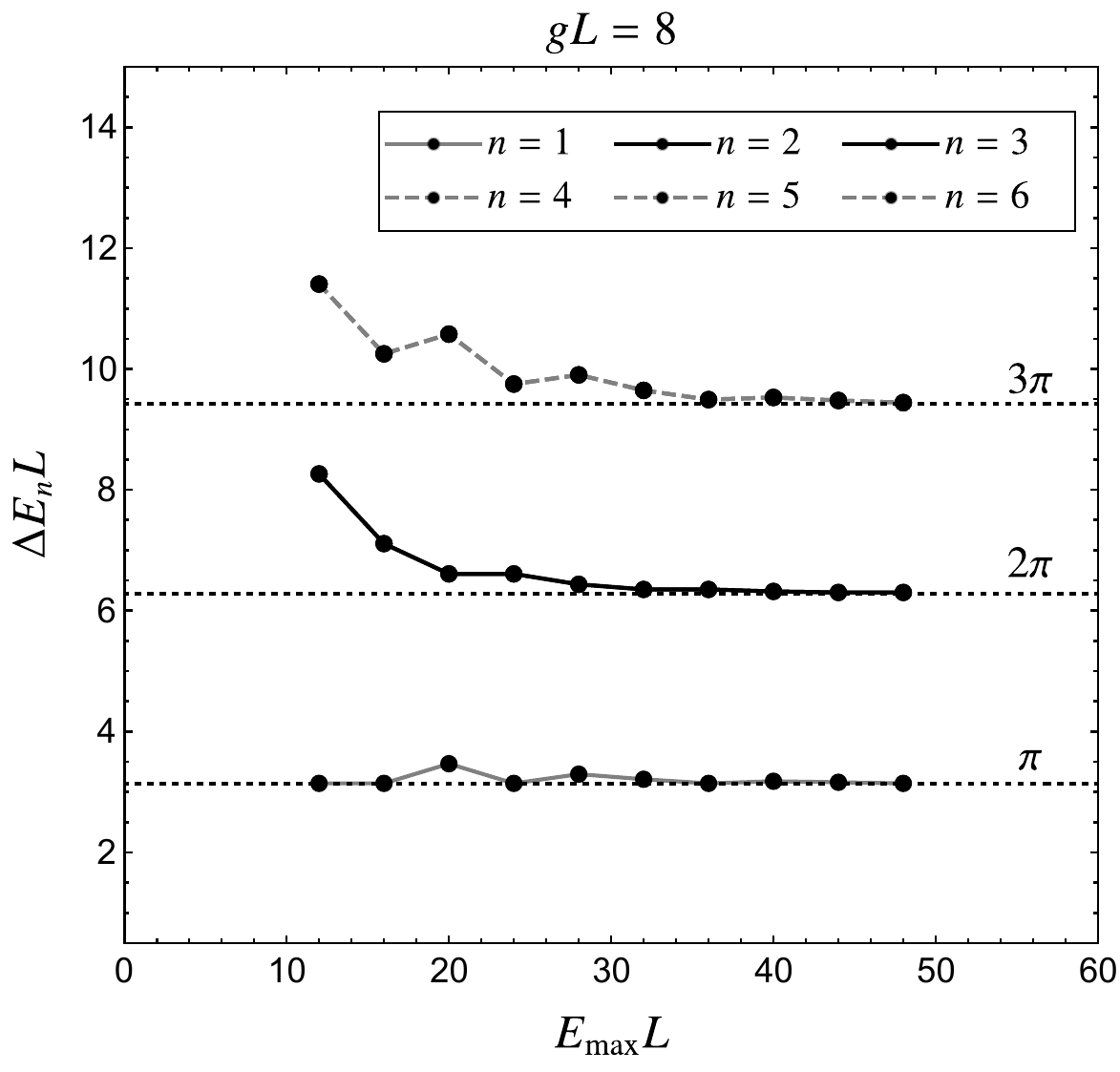} 
		\hspace{.001\textwidth}
		\includegraphics[height=0.46\columnwidth]{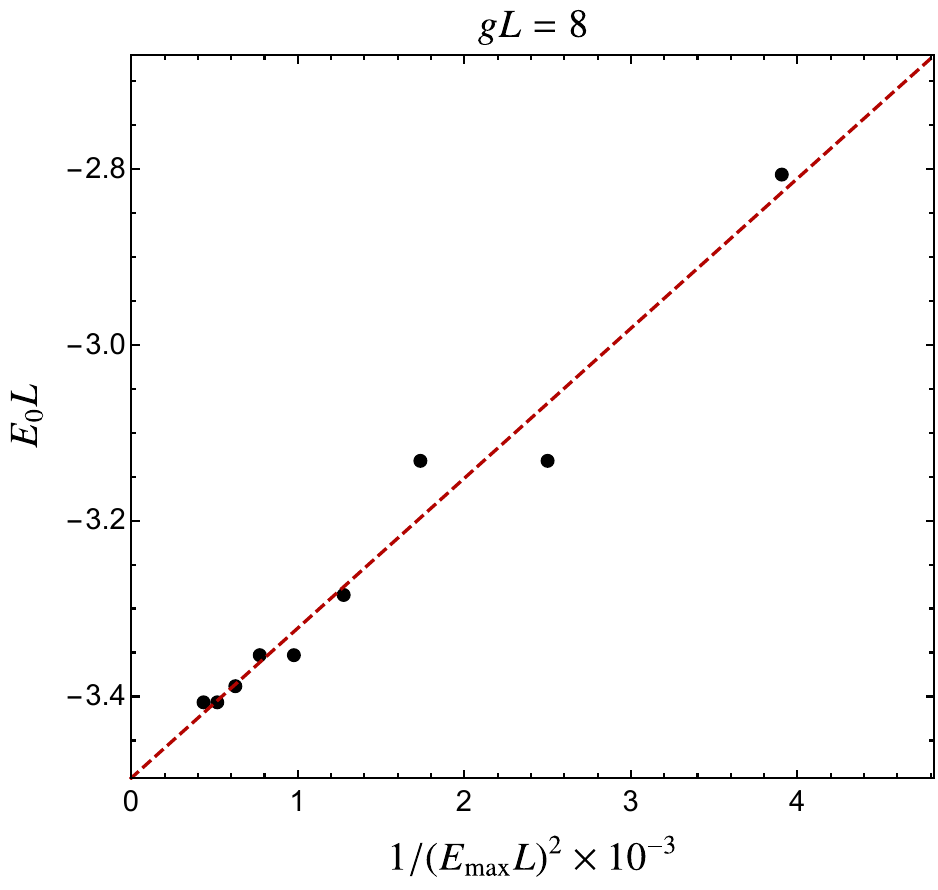}
		\caption{
			The left plot shows the excited energy gaps $\Delta E_n L$ of the $\SU(3)$ model  at moderate coupling $gL = 8$. The result from truncation is denoted by dots and connected by solid or dashed lines.  The first two energy gaps of the free theory are shown in dotted black lines.  The right plot shows vacuum energy $E_0 L$ plotted against $1/(\Emax L)^2$ in black dots, along with its linear fit given by a dashed red line.
}
	\label{fig:su3g3}
	\end{minipage}
\end{figure}

Our results for two fixed $g$ benchmarks are shown in Figs.~\ref{fig:su3g1} and~\ref{fig:su3g3}. We begin with $\SU(3)$ gauge theory at weak coupling. The right plot of Fig.~\ref{fig:su3g1} shows good agreement between the interacting theory and the free theory at weak gauge coupling. We expect a multiplet of three degenerate states with energies $\pi/L$. These are quark-antiquark states with vanishing diagonal nonabelian charges. Their color wavefunctions are linear combinations of $r\bar r,\ g\bar g,\ b\bar b$ states. The weak interaction lifts the energies of two states above the color singlet combination, i.e. $(r\bar r + g\bar g+ b\bar b)/\sqrt3$, for which both diagonal and off-diagonal nonabelian charges vanish. The left plot shows that the spectrum converges as expected with $1/(\Emax L)^2$. 



The $\SU(N)$ gauge theory with one Dirac fermion in $1+1$ dimensions exhibits gapless confinement, which is signaled in the infinite volume limit by the emergence of massless color-singlet bound states. We expect this because, in the infrared limit, this theory reduces to the $U(1)_N$ WZW conformal field theory, which is none other than the free compact scalar~\cite{Delmastro:2021otj}. As $gL\rightarrow\infty$, we expect the low energy spectrum we obtain from Hamiltonian truncation to match that of a free massless scalar on an interval. 

Demanding that the scalar field vanishes at the interval endpoints (as in the Schwinger model case), the oscillator modes of the free massless scalar have energies $j\pi/L$, for positive integer $j$. A general excited state therefore has an energy gap also given by Eq.~(\ref{eq:int-boson}), but with the Schwinger mass $M^2_S$ set to zero. Consequently, every energy level lies at $E=n\pi/L$ for some non-negative integer $n$. The degeneracy of the level at $n\pi/L$ is the number of integer partitions of $n$: each set of occupation numbers satisfying $\sum_j j n_j=n$ corresponds to a distinct way of writing $n$ as a sum of positive integers. In particular, the first excited state is a meson with energy $\pi/L$ above the ground state, while the higher levels exhibit increasingly large degeneracies.

We first test these expectations at moderate coupling in Fig.~\ref{fig:su3g3}. As shown in the left panel, color-singlet bound states emerge in the energy gap spectrum at the expected integer multiples of $\pi/L$, with the degeneracy pattern predicted by the free-scalar description. Remarkably, the agreement between the Hamiltonian-truncation data and these theoretical predictions is already extremely precise, despite the comparatively modest values of $E_{\rm max}$ accessible in our calculations. Figs.~\ref{fig:su3g1} and \ref{fig:su3g3} also provide clear qualitative evidence for color confinement: as the coupling is increased, states carrying nonabelian charge are lifted out of the low-energy spectrum, leaving behind states for which all nonabelian charges vanish.

\begin{table}[h!]
\centering
	\begin{minipage}{.9\textwidth}
	\centering
	\begin{tabular}{|c|c|}
	\hline
	$\Emax L$	& \# of states in basis	\\
	\hline
	16		&	194	\\
	32		& 	5,773	\\
	48		& 	88,873	\\
	\hline
	\end{tabular}
	\caption{The number of states in the Hilbert space of the nonabelian $\SU(3)$ model on the interval for a few benchmark values of $\Emax L$.}
	\label{tab:su3-basis}
	\end{minipage}
\end{table}

\begin{figure}[h!]
	\centering
	\begin{minipage}{.9\textwidth}{
	\centering
		\includegraphics[height=0.55\columnwidth]{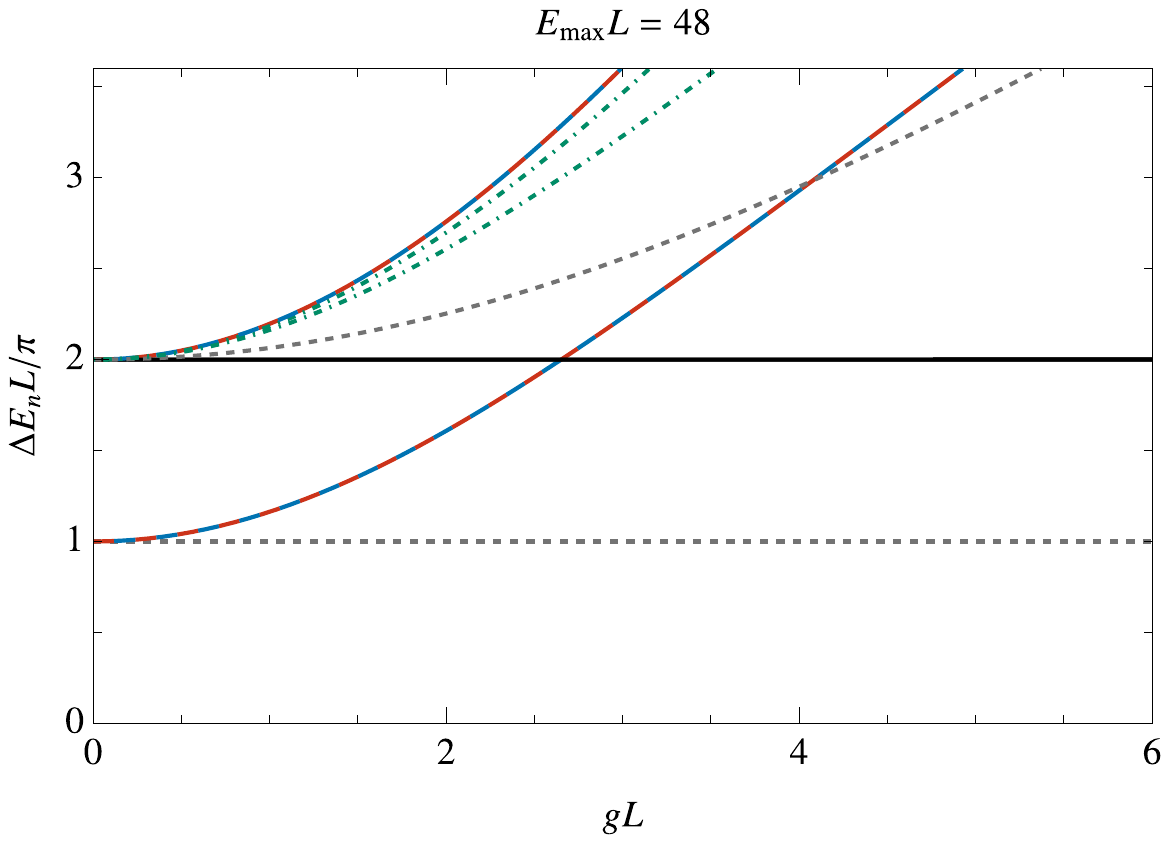} 
				\caption{This plot shows the 10 lowest excited energy gaps $\Delta E_n L$ of the $\SU(3)$ model as $gL$ is varied, using $E_{\rm max}L=48$. States which carry $\SU(3)$ color are shown by the red-and-blue dashed lines and the green dot-dashed lines, while states which are annihilated by all nonabelian charge operators, as defined in Eq.~(\ref{eq:nonabQ}), are shown by the gray dashed and black solid lines. Each red-and-blue line represents a pair of exactly degenerate states of different colors. The level represented by the black line is also doubly degenerate, while those represented by the dashed gray lines are nondegenerate.}
}
	\label{fig:su3-decouple}
	\end{minipage}
\end{figure}

In Fig.~\ref{fig:su3-decouple}, we show how the spectrum of low-lying energy states changes as the coupling is increased from weak to moderate strength for fixed $\Emax L=48$. This is the largest $\Emax L$ displayed in Fig.~\ref{fig:su3g3}, from which we conclude that the spectrum is well converged for the full range of couplings considered in Fig.~\ref{fig:su3-decouple}. A few interesting features emerge. First, we find that states which are nonsinglet under $\SU(3)$ color are lifted from the low energy spectrum\footnote{Color nonsinglet states can be identified by taking their corresponding eigenvectors, $\ket{\psi}$: if any charge operator $\mathcal{Q}^a$ defined in Eq.~(\ref{eq:nonabQ}) acts nontrivially, i.e. $\mathcal{Q}^a\ket{\psi}\neq0$, the state is not a color singlet.}. This is consistent with expectations that only confined color-singlet states should remain in the low energy spectrum in the infinite-coupling limit. The lowest-lying color singlet excitations track their asymptotic gaps of $n\pi/L$ throughout the modest range of $gL$ explored in Fig.~\ref{fig:su3-decouple}.

The upward-curving dashed gray level in Fig.~\ref{fig:su3-decouple} is qualitatively distinct from the nearly horizontal levels associated with the massless compact scalar: although it is also a color singlet, its energy increases with $gL$. We therefore interpret this state as a massive mesonic excitation. A quantitative determination of its mass, or a more detailed characterization of its properties, would require access to stronger couplings and larger truncated bases than are presently available.

More generally, going beyond these qualitative observations and quantitatively exploring the spectrum of excited meson states would require calculations at stronger coupling. At fixed $\Emax$, however, truncation errors grow rapidly with $g$, making it necessary to retain substantially more states in the truncated basis. The resulting increase in computational cost places such a study beyond the scope of the present work. We plan to investigate the approach to the infinite volume, $gL\rightarrow\infty$, limit of the nonabelian theory more systematically in future work.


\section{Discussion}
\label{sec:disc}

This work establishes Hamiltonian truncation as a practical nonperturbative tool for studying gauge theories in two dimensions, in equal time quantization. By developing explicit Hamiltonians for both abelian and nonabelian models on the spatial interval, and by numerically computing their spectra in a truncated low energy Hilbert space, we provide a first demonstration that these theories can be solved numerically, without ever needing to introduce a spatial lattice.

We began with quantum electrodynamics in 1+1D, the Schwinger model, where exact results from bosonization are available for comparison. Our Hamiltonian, given in Eq.~(\ref{eq:VSchwinger}), reproduces the known spectrum with high accuracy across a wide range of couplings, including the strongly coupled regime. This agreement validates the construction of the interacting Hamiltonian and the truncation procedure used to define the finite Hilbert space.

Building on this foundation, we derived the generalization of the Hamiltonian to nonabelian gauge groups, Eq.~(\ref{eq:VsuN}), and applied it to $\SU(3)$ gauge theory with a single massless Dirac fermion. The nonabelian formulation requires an additional fermion field for each color, leading to a much larger low–energy Hilbert space and an exponential increase in computational cost with $N$. Despite this extra complexity, the method offers a viable alternative for moderate $N$ and in practice yields a well-converged low–energy spectrum for $N=3$.

The $\SU(3)$ results contain signatures of color confinement: Color–nonsinglet states are included in the Hilbert space and appear in the low–energy spectrum when the gauge coupling is small (in units of the interval length), but disappear as the coupling grows and the confinement length falls well below the size of the interval. 

Within the baryon-number-zero sector at moderate coupling, where no exact solution exists, we nevertheless observe behaviour consistent with the expected deep-infrared limit in which the theory flows to a free compact scalar~\cite{Delmastro:2021otj}. In particular, the low-energy spectrum contains the full set of scalar-field levels at the expected integer multiples of $\pi/L$, with the predicted degeneracies reproduced to high precision. Extending the calculation to stronger couplings and larger basis sizes will be necessary to study the sector of massive mesonic excitations more quantitatively.

Working on the spatial interval brings both costs and benefits. Translational invariance is necessarily broken, and the fermion boundary conditions at the endpoints break chiral symmetry. These features complicate the interpretation of certain observables. There is, however, a benefit: all gauge degrees of freedom can be removed exactly, and the resulting formulation generalizes to nonabelian gauge groups with minimal additional structure. Now that the method has been established in this simple geometry, it provides a controlled starting point for future extensions to more elaborate setups that preserve a larger set of spacetime symmetries. 

We adopt axial gauge to achieve this, accepting the loss of manifest gauge invariance in exchange for a simple Hamiltonian. Truncating the Hilbert space in the free-theory eigenbasis introduces violations of gauge symmetry, but in two dimensions the strong relevance of gauge interactions ensures that these effects vanish rapidly with the energy cutoff. The excellent convergence of our numerical spectra provides direct evidence for this mechanism.

Having established and validated explicit Hamiltonians for abelian and nonabelian gauge theories in 1+1D on the interval, there are many natural directions for further work. The same formalism can be used both to study a wider range of theories, including those with nonzero mass, $\theta$ parameter, different gauge groups, or with multiple fermion flavors, and to compute new observables such as condensates, vacuum expectation values, and scattering amplitudes from real time dynamics. The methods developed here also offer a benchmark for alternative computational approaches to strongly coupled gauge theories.

\section*{Note added}
In the process of addressing reviewer comments, we discovered a bug in our code for generating the Hamiltonian of the $\SU(3)$ gauge theory. This has been corrected, and changes the results in Figs.~\ref{fig:su3g1}, \ref{fig:su3g3}, and \ref{fig:su3-decouple}. The logic of the analysis is unchanged, but the results now reflect the output of the corrected code.

\begin{acknowledgments}
We would like to thank Markus Luty, Tim Cohen and Kara Farnsworth for many helpful comments, suggestions and feedback on our draft. We would also like to thank Joan Elias Mir\'o and Soo-Jong Rey for useful discussions. We are grateful to the Mainz Institute
for Theoretical Physics (MITP) of the DFG Cluster of Excellence PRISMA+ (Project ID 39083149) for its hospitality and support during the initial stages of this work.
\end{acknowledgments}
\vspace{1.0cm}
\appendix

\section{Tensors Used in Building the Interaction}
\label{app:tensors}
It is convenient to express the coefficients from the interaction in Eq.~(\ref{eq:intschem}) in terms of the following function
\begin{align}
	f(A, B) = \begin{cases}
		\frac{1}{3}, & \text{if } A = B = 0 \\
		\frac{(-1)^{B+1}}{\pi^2 B^2}, & \text{if } A = 0,\, B\neq0 \\
		\frac{(-1)^{A+1}}{\pi^2 A^2}, & \text{if } B = 0,\, A\neq0 \\
		\frac{1}{2\pi^2B^2}, & \text{if } A=\pm B,\, A\neq 0 \\
		0 & \text{otherwise,}
	\end{cases}
	\label{eq:fcases}
\end{align}
which is defined for integer values of $A$ and $B$, and is symmetric in its two arguments. 

If we specialize to the massless $m=0$ case, the coefficients $V^{\alpha_1,\alpha_2,\alpha_3,\alpha_4}_{nmkl}$, which are defined as integrals in Eq.~(\ref{eq:vcoeffs1}), can be evaluated using
\begin{align}
	V^{ffff}_{nmkl} = V^{ff\bar{f}\bar{f}}_{nmkl} = V^{\bar{f}\bar{f}ff}_{nmkl} = V^{\bar{f}\bar{f}\bar{f}\bar{f}}_{nmkl} & = f(n-m,k-l)\,, \nonumber \\
	V^{fff\bar{f}}_{nmkl} = V^{ff\bar{f}f}_{nmkl} = V^{\bar{f}\bar{f}f\bar{f}}_{nmkl} = V^{\bar{f}\bar{f}\bar{f}f}_{nmkl}  & = f(n-m,k+l+1)\,, \nonumber \\
	V^{f\bar{f}\bar{f}\bar{f}}_{nmkl} = V^{\bar{f}f\bar{f}\bar{f}}_{nmkl} = V^{f\bar{f}ff}_{nmkl} = V^{\bar{f}fff}_{nmkl}& = f(n+m+1,k-l)\,, \nonumber \\
	V^{f\bar{f}f\bar{f}}_{nmkl}  = V^{f\bar{f}\bar{f}f}_{nmkl} = V^{\bar{f}ff\bar{f}}_{nmkl} = V^{\bar{f}f\bar{f}f}_{nmkl}     & = f(n+m+1,k+l+1)\,.
	\label{eq:vtensors}
\end{align}

The $\kappa^{(i)}$ tensors, which appear in Eqs.~(\ref{eq:VSchwinger}) and (\ref{eq:VsuN}), can be determined by normal ordering the terms in Eq.~(\ref{eq:intschem}) and inputting the expressions from Eq.~(\ref{eq:vtensors}). The explicit formulae for the tensors with four indices are
\begin{align}
	\kappa_{nmkl}^{(1)} & = - f(n+k+1, m+l+1)\,,\nonumber\\
	\kappa_{nmkl}^{(2)} &= 2f(n-l, m+k+1)\,,\nonumber\\
	\kappa_{nmkl}^{(3)}	&= - \frac12 f(n-k, m-l)\,,\nonumber\\
	\kappa_{nmkl}^{(4a)}	&= f(n+m+1, k+l+1)\,,\nonumber\\
	\kappa_{nmkl}^{(4b)}	&= -f(n-l, m-k)\,,
\end{align}
where for the Schwinger model, $\kappa^{(4)}_{nmkl} = \kappa^{(4a)}_{nmkl} + \kappa^{(4b)}_{nmkl}$, but for nonabelian gauge theory, they enter the interaction $V$, shown in Eq.~(\ref{eq:VsuN}), as independent tensors.

The tensors with two indices are given by
\begin{align}
	\kappa_{nl}^{(5)}		&= \sum_{m=0}^\infty
	f(n-m, m+l+1)
	- f(m-l, m+n+1)\,,
	\nonumber	\\
	\kappa_{nl}^{(6)}		&= \frac12 \sum_{m=0}^\infty 
	f(n-m, m-l)
	- f(n+m+1, m+l+1)\,.
\end{align}
These tensors contain infinite sums. They can be reduced to the following expressions, which do not contain infinite sums and which we use for our numerical analysis
\begin{align}
	\kappa_{nl}^{(5)}		&= \begin{cases} 
		\frac{1-(-1)^{n+l}}{\pi^2(n+l+1)^2} \,, & \text{if }  n>l \\
		-\frac{1-(-1)^{n+l}}{\pi^2(n+l+1)^2} \,, & \text{if }  l> n \\
		0\,, & \text{otherwise}
	\end{cases}  \nonumber\\
	\kappa_{nl}^{(6)}		&= \begin{cases}
		\frac{1}{6} + \sum_{k=1}^n \frac{1}{2\pi^2 k^2}\,,& \text{if }	n-l = 0 \\
		\frac{1}{\pi^2(n-l)^2}\,, & \text{if } n-l = \text{odd} \\
		0\,, & \text{otherwise.}
	\end{cases}
\end{align}

\bibliography{qcd-int}

@article{Ingoldby:2025bdb,
    author = "Ingoldby, James and Spannowsky, Michael and Sypchenko, Timur and Williams, Simon and Wingate, Matthew",
    title = "{Real-Time Scattering on Quantum Computers via Hamiltonian Truncation}",
    eprint = "2505.03878",
    archivePrefix = "arXiv",
    primaryClass = "quant-ph",
    reportNumber = "IPPP/25/24",
    month = "5",
    year = "2025"
}

@article{Demiray:2025zqh,
    author = "Demiray, Ekrem and Farnsworth, Kara and Houtz, Rachel",
    title = "{Systematic Improvement of Hamiltonian Truncation Effective Theory}",
    eprint = "2507.15941",
    archivePrefix = "arXiv",
    primaryClass = "hep-th",
    month = "7",
    year = "2025"
}

@article{Delmastro:2021otj,
	author = "Delmastro, Diego and Gomis, Jaume and Yu, Matthew",
	title = "{Infrared phases of 2d QCD}",
	eprint = "2108.02202",
	archivePrefix = "arXiv",
	primaryClass = "hep-th",
	doi = "10.1007/JHEP02(2023)157",
	journal = "JHEP",
	volume = "02",
	pages = "157",
	year = "2023"
}

@article{Ingoldby:2024fcy,
    author = "Ingoldby, James and Spannowsky, Michael and Sypchenko, Timur and Williams, Simon",
    title = "{Enhancing quantum field theory simulations on NISQ devices with Hamiltonian truncation}",
    eprint = "2407.19022",
    archivePrefix = "arXiv",
    primaryClass = "quant-ph",
    reportNumber = "IPPP/24/37",
    doi = "10.1103/PhysRevD.110.096016",
    journal = "Phys. Rev. D",
    volume = "110",
    number = "9",
    pages = "096016",
    year = "2024"
}

@article{Schmoll:2023eez,
    author = "Schmoll, Philipp and Naumann, Jan and Nietner, Alexander and Eisert, Jens and Sotiriadis, Spyros",
    title = "{Hamiltonian truncation tensor networks for quantum field theories}",
    eprint = "2312.12506",
    archivePrefix = "arXiv",
    primaryClass = "quant-ph",
    month = "12",
    year = "2023"
}

@article{Chen:2023glf,
    author = "Chen, Hongbin and Fitzpatrick, A. Liam and Katz, Emanuel and Xin, Yuan",
    title = "{Large momentum EFT and lightcone quantization}",
    eprint = "2306.13171",
    archivePrefix = "arXiv",
    primaryClass = "hep-th",
    doi = "10.1007/JHEP07(2025)008",
    journal = "JHEP",
    volume = "07",
    pages = "008",
    year = "2025"
}

@article{Ciavarella:2023mfc,
	author = "Ciavarella, Anthony N.",
	title = "{Quantum simulation of lattice QCD with improved Hamiltonians}",
	eprint = "2307.05593",
	archivePrefix = "arXiv",
	primaryClass = "hep-lat",
	reportNumber = "IQuS@UW-21-056",
	doi = "10.1103/PhysRevD.108.094513",
	journal = "Phys. Rev. D",
	volume = "108",
	number = "9",
	pages = "094513",
	year = "2023"
}

@article{Delouche:2023wsl,
	author = "Delouche, Olivier and Elias Miro, Joan and Ingoldby, James",
	title = "{Hamiltonian truncation crafted for UV-divergent QFTs}",
	eprint = "2312.09221",
	archivePrefix = "arXiv",
	primaryClass = "hep-th",
	reportNumber = "IPPP/23/79",
	doi = "10.21468/SciPostPhys.16.4.105",
	journal = "SciPost Phys.",
	volume = "16",
	number = "4",
	pages = "105",
	year = "2024"
}

@article{Fitzpatrick:2022dwq,
	author = "Fitzpatrick, A. Liam and Katz, Emanuel",
	title = "{Snowmass White Paper: Hamiltonian Truncation}",
	eprint = "2201.11696",
	archivePrefix = "arXiv",
	primaryClass = "hep-th",
	month = "1",
	year = "2022"
}

@article{EliasMiro:2022pua,
    author = "Elias Miro, Joan and Ingoldby, James",
    title = "{Effective Hamiltonians and Counterterms for Hamiltonian Truncation}",
    eprint = "2212.07266",
    archivePrefix = "arXiv",
    primaryClass = "hep-th",
    doi = "10.1007/JHEP07(2023)052",
    journal = "JHEP",
    volume = "07",
    pages = "052",
    year = "2023"
}

@article{Henning:2022xlj,
    author = "Henning, Brian and Murayama, Hitoshi and Riva, Francesco and Thompson, Jedidiah O. and Walters, Matthew T.",
    title = "{Towards a nonperturbative construction of the S-matrix}",
    eprint = "2209.14306",
    archivePrefix = "arXiv",
    primaryClass = "hep-th",
    doi = "10.1007/JHEP05(2023)197",
    journal = "JHEP",
    volume = "05",
    pages = "197",
    year = "2023"
}

@article{EliasMiro:2021aof,
    author = "Elias Miro, Joan and Ingoldby, James",
    title = "{Hamiltonian Truncation with larger dimensions}",
    eprint = "2112.09049",
    archivePrefix = "arXiv",
    primaryClass = "hep-th",
    doi = "10.1007/JHEP05(2022)151",
    journal = "JHEP",
    volume = "05",
    pages = "151",
    year = "2022"
}

@article{Cohen:2021erm,
	author = "Cohen, Timothy and Farnsworth, Kara and Houtz, Rachel and Luty, Markus A.",
	title = "{Hamiltonian Truncation Effective Theory}",
	eprint = "2110.08273",
	archivePrefix = "arXiv",
	primaryClass = "hep-th",
	doi = "10.21468/SciPostPhys.13.2.011",
	journal = "SciPost Phys.",
	volume = "13",
	number = "2",
	pages = "011",
	year = "2022"
}

@article{Anand:2020qnp,
    author = "Anand, Nikhil and Katz, Emanuel and Khandker, Zuhair U. and Walters, Matthew T.",
    title = "{Nonperturbative dynamics of (2+1)d $\phi^4$-theory from Hamiltonian truncation}",
    eprint = "2010.09730",
    archivePrefix = "arXiv",
    primaryClass = "hep-th",
    doi = "10.1007/JHEP05(2021)190",
    journal = "JHEP",
    volume = "05",
    pages = "190",
    year = "2021"
}

@article{Elias-Miro:2020qwz,
    author = "Elias-Mir{\'o}, Joan and Hardy, Edward",
    title = "{Exploring Hamiltonian Truncation in $\bf{d=2+1}$}",
    eprint = "2003.08405",
    archivePrefix = "arXiv",
    primaryClass = "hep-th",
    doi = "10.1103/PhysRevD.102.065001",
    journal = "Phys. Rev. D",
    volume = "102",
    number = "6",
    pages = "065001",
    year = "2020"
}

@article{James:2017cpc,
    author = "James, Andrew J. A. and Konik, Robert M. and Lecheminant, Philippe and Robinson, Neil J. and Tsvelik, Alexei M.",
    title = "{Non-perturbative methodologies for low-dimensional strongly-correlated systems: From non-abelian bosonization to truncated spectrum methods}",
    eprint = "1703.08421",
    archivePrefix = "arXiv",
    primaryClass = "cond-mat.str-el",
    doi = "10.1088/1361-6633/aa91ea",
    journal = "Rept. Prog. Phys.",
    volume = "81",
    number = "4",
    pages = "046002",
    year = "2018"
}

@article{Elias-Miro:2017tup,
    author = "Elias-Miro, Joan and Rychkov, Slava and Vitale, Lorenzo G.",
    title = "{NLO Renormalization in the Hamiltonian Truncation}",
    eprint = "1706.09929",
    archivePrefix = "arXiv",
    primaryClass = "hep-th",
    reportNumber = "CERN-TH-2017-124",
    doi = "10.1103/PhysRevD.96.065024",
    journal = "Phys. Rev. D",
    volume = "96",
    number = "6",
    pages = "065024",
    year = "2017"
}

@article{Anand:2017yij,
    author = "Anand, Nikhil and Genest, Vincent X. and Katz, Emanuel and Khandker, Zuhair U. and Walters, Matthew T.",
    title = "{RG flow from $\phi^4$ theory to the 2D Ising model}",
    eprint = "1704.04500",
    archivePrefix = "arXiv",
    primaryClass = "hep-th",
    doi = "10.1007/JHEP08(2017)056",
    journal = "JHEP",
    volume = "08",
    pages = "056",
    year = "2017"
}

@article{Bajnok:2015bgw,
    author = "Bajnok, Zoltan and Lajer, Marton",
    title = "{Truncated Hilbert space approach to the 2d $\phi^{4}$ theory}",
    eprint = "1512.06901",
    archivePrefix = "arXiv",
    primaryClass = "hep-th",
    doi = "10.1007/JHEP10(2016)050",
    journal = "JHEP",
    volume = "10",
    pages = "050",
    year = "2016"
}

@article{Elias-Miro:2015bqk,
    author = "Elias-Miro, J. and Montull, M. and Riembau, M.",
    title = "{The renormalized Hamiltonian truncation method in the large $E_T$ expansion}",
    eprint = "1512.05746",
    archivePrefix = "arXiv",
    primaryClass = "hep-th",
    doi = "10.1007/JHEP04(2016)144",
    journal = "JHEP",
    volume = "04",
    pages = "144",
    year = "2016"
}

@article{Rychkov:2015vap,
    author = "Rychkov, Slava and Vitale, Lorenzo G.",
    title = "{Hamiltonian truncation study of the $\phi^4$ theory in two dimensions. II. The $\mathbb Z_2$ -broken phase and the Chang duality}",
    eprint = "1512.00493",
    archivePrefix = "arXiv",
    primaryClass = "hep-th",
    reportNumber = "CERN-PH-TH-2015-277",
    doi = "10.1103/PhysRevD.93.065014",
    journal = "Phys. Rev. D",
    volume = "93",
    number = "6",
    pages = "065014",
    year = "2016"
}

@article{Rychkov:2014eea,
    author = "Rychkov, Slava and Vitale, Lorenzo G.",
    title = "{Hamiltonian truncation study of the $?^4$ theory in two dimensions}",
    eprint = "1412.3460",
    archivePrefix = "arXiv",
    primaryClass = "hep-th",
    reportNumber = "CERN-PH-TH-2014-254",
    doi = "10.1103/PhysRevD.91.085011",
    journal = "Phys. Rev. D",
    volume = "91",
    pages = "085011",
    year = "2015"
}

@article{Hogervorst:2014rta,
    author = "Hogervorst, Matthijs and Rychkov, Slava and van Rees, Balt C.",
    title = "{Truncated conformal space approach in d dimensions: A cheap alternative to lattice field theory?}",
    eprint = "1409.1581",
    archivePrefix = "arXiv",
    primaryClass = "hep-th",
    reportNumber = "CERN-PH-TH-2014-155",
    doi = "10.1103/PhysRevD.91.025005",
    journal = "Phys. Rev. D",
    volume = "91",
    pages = "025005",
    year = "2015"
}

@article{Katz:2014uoa,
    author = "Katz, Emanuel and Marques Tavares, Gustavo and Xu, Yiming",
    title = "{A solution of 2D QCD at Finite $N$ using a conformal basis}",
    eprint = "1405.6727",
    archivePrefix = "arXiv",
    primaryClass = "hep-th",
    month = "5",
    year = "2014"
}

@article{Yurov:1991my,
    author = "Yurov, V. P. and Zamolodchikov, Alexei B.",
    title = "{Truncated fermionic space approach to the critical 2-D Ising model with magnetic field}",
    doi = "10.1142/S0217751X91002161",
    journal = "Int. J. Mod. Phys. A",
    volume = "6",
    pages = "4557--4578",
    year = "1991"
}

@article{Yurov:1989yu,
    author = "Yurov, V. P. and Zamolodchikov, A. B.",
    title = "{TRUNCATED CONFORMAL SPACE APPROACH TO SCALING LEE-YANG MODEL}",
    reportNumber = "ITEP-89-161",
    doi = "10.1142/S0217751X9000218X",
    journal = "Int. J. Mod. Phys. A",
    volume = "5",
    pages = "3221--3246",
    year = "1990"
}

@article{Brooks:1983sb,
    author = "Brooks, III, E. D. and Frautschi, Steven C.",
    title = "{Scalars Coupled to Fermions in (1+1)-dimensions}",
    reportNumber = "CALT-68-1069",
    doi = "10.1007/BF01546194",
    journal = "Z. Phys. C",
    volume = "23",
    pages = "263",
    year = "1984"
}

@article{Nielsen:1980rz,
    author = "Nielsen, Holger Bech and Ninomiya, M.",
    editor = "Julve, J. and Ram{\'o}n-Medrano, M.",
    title = "{Absence of Neutrinos on a Lattice. 1. Proof by Homotopy Theory}",
    reportNumber = "RL-80-090",
    doi = "10.1016/0550-3213(82)90011-6",
    journal = "Nucl. Phys. B",
    volume = "185",
    pages = "20",
    year = "1981",
    note = "[Erratum: Nucl.Phys.B 195, 541 (1982)]"
}

@article{Coleman:1975pw,
    author = "Coleman, Sidney R. and Jackiw, R. and Susskind, Leonard",
    title = "{Charge Shielding and Quark Confinement in the Massive Schwinger Model}",
    reportNumber = "MIT-CTP-470",
    doi = "10.1016/0003-4916(75)90212-2",
    journal = "Annals Phys.",
    volume = "93",
    pages = "267",
    year = "1975"
}

@article{Wilson:1974sk,
    author = "Wilson, Kenneth G.",
    editor = "Taylor, J. C.",
    title = "{Confinement of Quarks}",
    reportNumber = "CLNS-262",
    doi = "10.1103/PhysRevD.10.2445",
    journal = "Phys. Rev. D",
    volume = "10",
    pages = "2445--2459",
    year = "1974"
}

@article{Schwinger:1962tp,
    author = "Schwinger, Julian S.",
    title = "{Gauge Invariance and Mass. 2.}",
    doi = "10.1103/PhysRev.128.2425",
    journal = "Phys. Rev.",
    volume = "128",
    pages = "2425--2429",
    year = "1962"
}

@article{coleman1976more,
	title={More about the massive Schwinger model},
	author={Coleman, Sidney},
	journal={Annals of Physics},
	volume={101},
	number={1},
	pages={239--267},
	year={1976},
	publisher={Elsevier}
}

@article{Okuda:2022hsq,
	author = "Okuda, Takuya",
	title = "{Schwinger model on an interval: Analytic results and DMRG}",
	eprint = "2210.00297",
	archivePrefix = "arXiv",
	primaryClass = "hep-lat",
	reportNumber = "UT-Komaba/22-4",
	doi = "10.1103/PhysRevD.107.054506",
	journal = "Phys. Rev. D",
	volume = "107",
	number = "5",
	pages = "054506",
	year = "2023"
}

@article{Anand:2021qnd,
	author = "Anand, Nikhil and Fitzpatrick, A. Liam and Katz, Emanuel and Xin, Yuan",
	title = "{Chiral limit of 2d QCD revisited with lightcone conformal truncation}",
	eprint = "2111.00021",
	archivePrefix = "arXiv",
	primaryClass = "hep-th",
	doi = "10.1007/JHEP01(2024)189",
	journal = "JHEP",
	volume = "01",
	pages = "189",
	year = "2024"
}

@article{PhysRev.127.1821,
	title = {Quantization of the Yang-Mills Field},
	author = {Arnowitt, R. L. and Fickler, S. I.},
	journal = {Phys. Rev.},
	volume = {127},
	issue = {5},
	pages = {1821--1829},
	numpages = {0},
	year = {1962},
	month = {Sep},
	publisher = {American Physical Society},
	doi = {10.1103/PhysRev.127.1821},
	url = {https://link.aps.org/doi/10.1103/PhysRev.127.1821}
}

@book{Weinberg:1995mt,
	author = "Weinberg, Steven",
	title = "{The Quantum theory of fields. Vol. 1: Foundations}",
	doi = "10.1017/CBO9781139644167",
	isbn = "978-0-521-67053-1, 978-0-511-25204-4",
	publisher = "Cambridge University Press",
	month = "6",
	year = "2005"
}

@article{Farrell:2022wyt,
	author = "Farrell, Roland C. and Chernyshev, Ivan A. and Powell, Sarah J. M. and Zemlevskiy, Nikita A. and Illa, Marc and Savage, Martin J.",
	title = "{Preparations for quantum simulations of quantum chromodynamics in 1+1 dimensions. I. Axial gauge}",
	eprint = "2207.01731",
	archivePrefix = "arXiv",
	primaryClass = "quant-ph",
	reportNumber = "IQuS@UW-21-027, NT@UW-22-05",
	doi = "10.1103/PhysRevD.107.054512",
	journal = "Phys. Rev. D",
	volume = "107",
	number = "5",
	pages = "054512",
	year = "2023"
}

@article{Anand:2020gnn,
	author = "Anand, Nikhil and Fitzpatrick, A. Liam and Katz, Emanuel and Khandker, Zuhair U. and Walters, Matthew T. and Xin, Yuan",
	title = "{Introduction to Lightcone Conformal Truncation: QFT Dynamics from CFT Data}",
	eprint = "2005.13544",
	archivePrefix = "arXiv",
	primaryClass = "hep-th",
	month = "5",
	year = "2020"
}

@article{Bhanot:1993xp,
    author = "Bhanot, Gyan and Demeterfi, Kresimir and Klebanov, Igor R.",
    title = "{(1+1)-dimensional large N QCD coupled to adjoint fermions}",
    eprint = "hep-th/9307111",
    archivePrefix = "arXiv",
    reportNumber = "PUPT-1413, IASSNS-HEP-93-42, NSF-ITP-93-107I",
    doi = "10.1103/PhysRevD.48.4980",
    journal = "Phys. Rev. D",
    volume = "48",
    pages = "4980--4990",
    year = "1993"
}

@article{Demeterfi:1993rs,
    author = "Demeterfi, Kresimir and Klebanov, Igor R. and Bhanot, Gyan",
    title = "{Glueball spectrum in a (1+1)-dimensional model for QCD}",
    eprint = "hep-th/9311015",
    archivePrefix = "arXiv",
    reportNumber = "PUPT-1427, IASSNS-HEP-93-59",
    doi = "10.1016/0550-3213(94)90236-4",
    journal = "Nucl. Phys. B",
    volume = "418",
    pages = "15--29",
    year = "1994"
}

@article{Gopakumar:2012gd,
    author = "Gopakumar, Rajesh and Hashimoto, Akikazu and Klebanov, Igor R. and Sachdev, Subir and Schoutens, Kareljan",
    title = "{Strange Metals in One Spatial Dimension}",
    eprint = "1206.4719",
    archivePrefix = "arXiv",
    primaryClass = "hep-th",
    reportNumber = "PUPT-2419, MAD-TH-12-06",
    doi = "10.1103/PhysRevD.86.066003",
    journal = "Phys. Rev. D",
    volume = "86",
    pages = "066003",
    year = "2012"
}

@article{Dempsey:2021xpf,
    author = "Dempsey, Ross and Klebanov, Igor R. and Pufu, Silviu S.",
    title = "{Exact symmetries and threshold states in two-dimensional models for QCD}",
    eprint = "2101.05432",
    archivePrefix = "arXiv",
    primaryClass = "hep-th",
    reportNumber = "PUPT-2623",
    doi = "10.1007/JHEP10(2021)096",
    journal = "JHEP",
    volume = "10",
    pages = "096",
    year = "2021"
}

@article{Katz:2013qua,
	author = "Katz, Emanuel and Marques Tavares, Gustavo and Xu, Yiming",
	title = "{Solving 2D QCD with an adjoint fermion analytically}",
	eprint = "1308.4980",
	archivePrefix = "arXiv",
	primaryClass = "hep-th",
	doi = "10.1007/JHEP05(2014)143",
	journal = "JHEP",
	volume = "05",
	pages = "143",
	year = "2014"
}

@article{Lunin:1999ib,
    author = "Lunin, O. and Pinsky, S.",
    editor = "Ji, C. R. and Min, D. P.",
    title = "{SDLCQ: Supersymmetric discrete light cone quantization}",
    eprint = "hep-th/9910222",
    archivePrefix = "arXiv",
    reportNumber = "OHSTPY-HEP-T-99-018",
    doi = "10.1063/1.1301663",
    journal = "AIP Conf. Proc.",
    volume = "494",
    number = "1",
    pages = "140--218",
    year = "1999"
}

@article{Fitzpatrick:2019cif,
    author = "Fitzpatrick, A. Liam and Katz, Emanuel and Walters, Matthew T. and Xin, Yuan",
    title = "{Solving the 2D SUSY Gross-Neveu-Yukawa model with conformal truncation}",
    eprint = "1911.10220",
    archivePrefix = "arXiv",
    primaryClass = "hep-th",
    doi = "10.1007/JHEP01(2021)182",
    journal = "JHEP",
    volume = "01",
    pages = "182",
    year = "2021"
}

@article{Fitzpatrick:2023aqm,
	author = "Fitzpatrick, Andrew Liam and Katz, Emanuel and Xin, Yuan",
	title = "{Lightcone Hamiltonian for Ising field theory I: $T < T_c$}",
	eprint = "2311.16290",
	archivePrefix = "arXiv",
	primaryClass = "hep-th",
	doi = "10.21468/SciPostPhys.18.6.179",
	journal = "SciPost Phys.",
	volume = "18",
	number = "6",
	pages = "179",
	year = "2025"
}

@article{Fitzpatrick:2025hqk,
	author = "Fitzpatrick, A. Liam and Novikova, Anastasiia and Ring, Noah",
	title = "{Hamiltonian Truncation of Large $N_f$ QED and Large $N$ Vector-like Theories in $d=2+1$}",
	eprint = "2507.22103",
	archivePrefix = "arXiv",
	primaryClass = "hep-th",
	month = "7",
	year = "2025"
}

@article{Liu:2020eoa,
	author = "Liu, Junyu and Xin, Yuan",
	title = "{Quantum simulation of quantum field theories as quantum chemistry}",
	eprint = "2004.13234",
	archivePrefix = "arXiv",
	primaryClass = "hep-th",
	reportNumber = "CALT-TH-2020-009",
	doi = "10.1007/JHEP12(2020)011",
	journal = "JHEP",
	volume = "12",
	pages = "011",
	year = "2020"
}

@article{Bergknoff:1976xr,
	author = "Bergknoff, Hugh",
	title = "{Physical Particles of the Massive Schwinger Model}",
	reportNumber = "CLNS-341",
	doi = "10.1016/0550-3213(77)90204-8",
	journal = "Nucl. Phys. B",
	volume = "122",
	pages = "215--236",
	year = "1977"
}

@article{Dempsey:2022nys,
author = "Dempsey, Ross and Klebanov, Igor R. and Pufu, Silviu S. and Zan, Bernardo",
title = "{Discrete chiral symmetry and mass shift in the lattice Hamiltonian approach to the Schwinger model}",
eprint = "2206.05308",
archivePrefix = "arXiv",
primaryClass = "hep-th",
reportNumber = "PUPT-2634",
doi = "10.1103/PhysRevResearch.4.043133",
journal = "Phys. Rev. Res.",
volume = "4",
number = "4",
pages = "043133",
year = "2022"
}

@article{Delouche:2024yuo,
author = "Delouche, Olivier and Elias Miro, Joan and Ingoldby, James",
title = "{Testing the RG-flow M(3, 10) + {\ensuremath{\phi}}$_{1,7}$ {\textrightarrow} M(3, 8) with Hamiltonian Truncation}",
eprint = "2412.09295",
archivePrefix = "arXiv",
primaryClass = "hep-th",
reportNumber = "IPPP/24/79",
doi = "10.1007/JHEP04(2025)144",
journal = "JHEP",
volume = "04",
pages = "144",
year = "2025"
}

@article{Banuls:2019rao,
	author = "Ba{\~n}uls, Mari Carmen and Cichy, Krzysztof",
	title = "{Review on Novel Methods for Lattice Gauge Theories}",
	eprint = "1910.00257",
	archivePrefix = "arXiv",
	primaryClass = "hep-lat",
	doi = "10.1088/1361-6633/ab6311",
	journal = "Rept. Prog. Phys.",
	volume = "83",
	number = "2",
	pages = "024401",
	year = "2020"
}

@article{Banuls:2019bmf,
	author = "Ba{\~n}uls, M. C. and others",
	title = "{Simulating Lattice Gauge Theories within Quantum Technologies}",
	eprint = "1911.00003",
	archivePrefix = "arXiv",
	primaryClass = "quant-ph",
	doi = "10.1140/epjd/e2020-100571-8",
	journal = "Eur. Phys. J. D",
	volume = "74",
	number = "8",
	pages = "165",
	year = "2020"
}

@article{Dempsey:2025wia,
	author = "Dempsey, Ross and Gl{\"u}ck, Anna-Maria E. and Pufu, Silviu S. and Sogaard, Benjamin T.",
	title = "{Infinite matrix product states for (1+1)-dimensional gauge theories}",
	eprint = "2508.16363",
	archivePrefix = "arXiv",
	primaryClass = "hep-th",
	reportNumber = "PUPT-2657",
	month = "8",
	year = "2025"
}

@article{Montangero:2021puw,
	author = "Montangero, Simone and Rico, Enrique and Silvi, Pietro",
	title = "{Loop-free tensor networks for high-energy physics}",
	eprint = "2109.11842",
	archivePrefix = "arXiv",
	primaryClass = "quant-ph",
	doi = "10.1098/rsta.2021.0065",
	journal = "Phil. Trans. A. Math. Phys. Eng. Sci.",
	volume = "380",
	number = "2216",
	pages = "20210065",
	year = "2021"
}

@article{Abel:2025pxa,
	author = "Abel, Steven and Spannowsky, Michael and Williams, Simon",
	title = "{Qumode Tensor Networks for False Vacuum Decay in Quantum Field Theory}",
	eprint = "2506.17388",
	archivePrefix = "arXiv",
	primaryClass = "quant-ph",
	reportNumber = "IPPP/25/37",
	month = "6",
	year = "2025"
}

@article{Halimeh:2025vvp,
	author = "Halimeh, Jad C. and Mueller, Niklas and Knolle, Johannes and Papi{\'c}, Zlatko and Davoudi, Zohreh",
	title = "{Quantum simulation of out-of-equilibrium dynamics in gauge theories}",
	eprint = "2509.03586",
	archivePrefix = "arXiv",
	primaryClass = "quant-ph",
	month = "9",
	year = "2025"
}

@article{Bauer:2022hpo,
	author = "Bauer, Christian W. and others",
	title = "{Quantum Simulation for High-Energy Physics}",
	eprint = "2204.03381",
	archivePrefix = "arXiv",
	primaryClass = "quant-ph",
	reportNumber = "UMD-PP-022-04, LA-UR-22-22100, RIKEN-iTHEMS-Report-22, RIKEN-iTHEMS-Report-22,
	FERMILAB-PUB-22-249-SQMS-T, IQuS@UW-21-027, MITRE-21-03848-2, FERMILAB-PUB-22-249-SQMS-T",
	doi = "10.1103/PRXQuantum.4.027001",
	journal = "PRX Quantum",
	volume = "4",
	number = "2",
	pages = "027001",
	year = "2023"
}

@article{DiMeglio:2023nsa,
	author = "Di Meglio, Alberto and others",
	title = "{Quantum Computing for High-Energy Physics: State of the Art and Challenges}",
	eprint = "2307.03236",
	archivePrefix = "arXiv",
	primaryClass = "quant-ph",
	reportNumber = "FERMILAB-PUB-23-468-ETD",
	doi = "10.1103/PRXQuantum.5.037001",
	journal = "PRX Quantum",
	volume = "5",
	number = "3",
	pages = "037001",
	year = "2024"
}

@article{Halimeh:2023lid,
	author = "Halimeh, Jad C. and Aidelsburger, Monika and Grusdt, Fabian and Hauke, Philipp and Yang, Bing",
	title = "{Cold-atom quantum simulators of gauge theories}",
	eprint = "2310.12201",
	archivePrefix = "arXiv",
	primaryClass = "cond-mat.quant-gas",
	doi = "10.1038/s41567-024-02721-8",
	journal = "Nature Phys.",
	volume = "21",
	number = "1",
	pages = "25--36",
	year = "2025"
}

@article{PhysRevLett.46.77,
	title = {Monte Carlo Calculation of the Thermodynamic Properties of a Quantum Model: A One-Dimensional Fermion Lattice Model},
	author = {De Raedt, Hans and Lagendijk, Ad},
	journal = {Phys. Rev. Lett.},
	volume = {46},
	issue = {2},
	pages = {77--80},
	numpages = {0},
	year = {1981},
	month = {Jan},
	publisher = {American Physical Society},
	doi = {10.1103/PhysRevLett.46.77},
	url = {https://link.aps.org/doi/10.1103/PhysRevLett.46.77}
}

@article{VONDERLINDEN199253,
	title = {A quantum Monte Carlo approach to many-body physics},
	journal = {Physics Reports}, 
	volume = {220},
	number = {2},
	pages = {53-162},
	year = {1992},
	issn = {0370-1573},
	doi = {https://doi.org/10.1016/0370-1573(92)90029-Y},
	url = {https://www.sciencedirect.com/science/article/pii/037015739290029Y},
	author = {Wolfgang {von der Linden}}
}

@article{PhysRevE.49.3855,
	title = {Monte Carlo simulations with indefinite and complex-valued measures},
	author = {Kieu, T. D. and Griffin, C. J.},
	journal = {Phys. Rev. E},
	volume = {49},
	issue = {5},
	pages = {3855--3859},
	numpages = {0},
	year = {1994},
	month = {May},
	publisher = {American Physical Society},
	doi = {10.1103/PhysRevE.49.3855},
	url = {https://link.aps.org/doi/10.1103/PhysRevE.49.3855}
}

@article{Fontana:2024rux,
     author = "Fontana, Pierpaolo and Riaza, Marc Miranda and Celi, Alessio",
     title = "{Efficient Finite-Resource Formulation of Non-Abelian Lattice Gauge Theories beyond One Dimension}",
     eprint = "2409.04441",
     archivePrefix = "arXiv",
     primaryClass = "quant-ph",
     doi = "10.1103/k9p6-c649",
     journal = "Phys. Rev. X",
     volume = "15",
     number = "3",
     pages = "031065",
     year = "2025"}
\bibliographystyle{utphys}


\end{document}